\documentclass[journal]{IEEEtran}

\usepackage[english]{babel}
\usepackage{calc}
\usepackage{amsmath,amssymb,amsfonts}
\usepackage{bbm}
\usepackage{xcolor}
\usepackage[all]{xy} 
\usepackage[noadjust]{cite}
\usepackage{array}
\usepackage[tight,footnotesize]{subfigure}
\usepackage{graphicx}
\graphicspath{{fig_pdf/}}
\DeclareGraphicsExtensions{.pdf,.png}

\usepackage{algorithm}
\usepackage[noend]{algpseudocode}
\usepackage{epstopdf}
\usepackage{url}
\usepackage{multirow}

\newcommand{\rpl}{\text{\sc rpl}}

\newcommand{\msminone}{\text{min1}}
\newcommand{\msmintwo}{\text{min2}}
\newcommand{\msidxone}{\text{indx}\_\text{min1}}
\newcommand{\m}{\mathfrak{m}}

\algrenewcommand\algorithmicforall{\hspace*{-.5\parindent}\textbf{for all}}
\algrenewcommand\algorithmicif{\hspace*{-.5\parindent}\textbf{if}}

\newlength{\EndIterLen}
\algblockdefx[Init]{Init}{EndInit}{\hspace*{-.4\parindent}\textbf{Initialization}}{}
\algblockdefx[Iter]{Iter}{EndIter}{\hspace*{-.4\parindent}\textbf{Iteration Loop}}%
{\hspace*{-.35\parindent}\textbf{End Iteration Loop}\settowidth{\EndIterLen}{\textbf{End Iteration Loop}}}
\algblockdefx[MyWhile]{MyWhile}{EndMyWhile}{\textbf{While}}%
{\textbf{End While}\settowidth{\EndIterLen}{\textbf{End Iteration Loop}}}
\algtext*{EndInit}

\newcommand{\vect}[1]{\underline{\boldsymbol #1}}

\newtheorem{theorem}{Theorem}
\newtheorem{definition}[theorem]{Definition}
\newtheorem{proposition}[theorem]{Proposition}

\begin{document}
%
\title{Analysis and Design of Cost-Effective, High-Throughput LDPC Decoders}
%
%
%

\author{Thien Truong Nguyen-Ly, Valentin Savin, Khoa Le, David Declercq, 
Fakhreddine Ghaffari, and Oana Boncalo
\thanks{Thien Truong Nguyen-Ly is with CEA-LETI, MINATEC Campus, Grenoble, France, and 
ETIS\;ENSEA\,/\,UCP\,/\,CNRS\;UMR-8051, Cergy-Pontoise, France (e-mail: thientruong.nguyen-ly@cea.fr).}
\thanks{Valentin Savin is with CEA-LETI, MINATEC Campus, Grenoble, France (e-mail: valentin.savin@cea.fr).}
\thanks{Khoa Le, David Declercq, and Fakhreddine Ghaffari are with ETIS\;ENSEA\,/\,UCP\,/\,CNRS\;UMR-8051, Cergy-Pontoise, France (e-mail: \{khoa.letrung, ghaffari, declercq\}@ensea.fr).}%
\thanks{Oana Boncalo is with the Computers and Information Technology Department of the University Politehnica Timisoara, Romania (e-mail: boncalo@cs.upt.ro).}}
\maketitle

\begin{abstract}
This paper introduces a new approach to cost-effective, high-throughput hardware designs for Low Density Parity Check (LDPC) decoders. The proposed approach, called Non-Surjective Finite Alphabet Iterative Decoders (NS-FAIDs), exploits the robustness of message-passing LDPC decoders to inaccuracies in the calculation of exchanged messages, and it is shown to provide a unified framework for several designs previously proposed in the literature. 
NS-FAIDs are optimized by density evolution for regular and irregular LDPC codes, and are shown to provide different trade-offs between hardware complexity and decoding performance. Two hardware architectures targeting high-throughput applications are also proposed, integrating both Min-Sum (MS) and NS-FAID decoding kernels. ASIC post synthesis implementation results on $65$nm CMOS technology show that NS-FAIDs yield significant improvements in the throughput to area ratio, by up to $58.75\%$ with respect to the MS decoder, with even better or only slightly degraded error correction performance.
\end{abstract}


%
\IEEEpeerreviewmaketitle

  
\section{Introduction}
%
%
%
%

\IEEEPARstart{T}{he} increasing demand of massive data rates in wireless communication systems will require significantly higher processing speed of the baseband signal, as compared to conventional solutions. This is especially challenging for Forward Error Correction (FEC) mechanisms, since FEC decoding is one of the most computationally intensive baseband processing tasks, consuming a large amount of hardware resources and energy. The use of very large bandwidths will also result in stringent, application-specific, requirements in terms of both throughput and latency. The conventional approach to increase throughput is to use massively parallel architectures. In this context, 
Low-Density Parity-Check (LDPC) codes are recognized as the foremost solution, due to the intrinsic capacity of their decoders to accommodate various degrees of parallelism. They have found extensive applications in modern communication systems, due to their excellent decoding performance, high throughput capabilities 
\cite{Marjan2004, chen2011memory, chandrasetty2015resource, Kai2009}, and power efficiency \cite{Xiao2011,xiang2011847}, and have been adopted in several recent communication standards.

This paper targets the design of cost-effective, high-throughput LDPC decoders. One important characteristic of LDPC decoders is that the memory and interconnect blocks dominate the overall area/delay/power performance of the hardware design \cite{boutillon2014hardware}. To address this issue, we build upon the concept of Finite Alphabet Iterative Decoders (FAIDs), introduced in \cite{planjery2010iterative, planjery2011finite, planjery2013finite}. While FAIDs have been previously investigated for variable-node regular LDPC codes over the binary symmetric channel, this paper extends their use to any channel model, and to both regular and irregular LDPC codes.

The approach considered in this paper, referred to as {\em Non-Surjective FAIDs} (NS-FAIDs), is to allow storing the exchanged messages using a lower precision (smaller number of bits) than that used by the processing units. 
The basic idea is to reduce the size of the exchanged messages, once they have been updated by the processing units. Hence, to some extent, the proposed approach is akin to the use of {\em imprecise storage}, which is seen as an enabler for cost and throughput optimizations. Moreover, NS-FAIDs are shown to provide a unified framework for several designs previously proposed in the literature, including the normalized and offset Min-Sum (MS) decoders \cite{chen2005reduced, savin2014ldpc}, the partially offset MS decoder \cite{nguyenly2015fpga}, the MS-based decoders proposed in \cite{oh2010min, chandrasetty2012area}, or the recently introduced dual-quantization domain MS decoder \cite{abu2014low}.

This paper refines and extends some of the concepts we previously introduced in \cite{nguyen2016non, nguyen2017high}. In particular, the definition of NS-FAIDs \cite{nguyen2016non} is extended such as to cover a larger class of decoders, which is shown to significantly improve the decoding performance in case that the exchanged messages are quantized on a small number of bits (e.g., $2$ bits per exchanged message). We show that NS-FAIDs can be optimized by using the Density Evolution (DE) technique, so as to obtain the best possible decoding performance for given hardware constraints, expressed in terms of memory size reduction. The DE optimization is illustrated for both regular and irregular  LDPC codes, for which we propose a number of NS-FAIDs with different trade-offs between hardware complexity and decoding performance. 

 To assess the benefits of the NS-FAID approach, we further extend the hardware architectures proposed in \cite{nguyen2017high} to cover the case of irregular codes, and provide implementation results targeting an ASIC technology, which is more likely to reflect the benefits of the proposed NS-FAID approach in terms of throughput/area trade-off.
The proposed architectures target high-throughput and efficient use of the hardware resources. Both architectures implement layered decoding with fully parallel processing units. The first architecture is pipelined, so as to increase throughput and ensure an efficient use of the hardware resources, which in turn imposes specific constraints on the decoding layers\footnote{A decoding layer may consist of one or several rows of the base matrix of the QC-LDPC code, assuming that they do not overlap.}, in order to ensure proper execution of the layered decoding process. The second architecture does not make use of pipelining, but allows maximum parallelism to be exploited through the use of  full decoding layers\footnote{A decoding layer is said to be full if each column of the base matrix has one non-negative entry in one of the rows composing the layer.}, thus resulting in significant increase in throughput. Both MS and NS-FAID decoding kernels are integrated into each of the two proposed architectures, and compared in terms of area and throughput. ASIC post synthesis implementation results on $65$nm CMOS technology show a throughput to area ratio improvement by up to $58.75\%$, when the NS-FAID kernel is used, with even better or only slightly degraded error correction performance.

The rest of the paper is organized as follows. NS-FAIDs are introduced in Section~\ref{sec:ns_faids}, which also discusses their expected implementation benefits and the DE analysis. The optimization of regular and irregular NS-FAIDs is presented in Section~\ref{sec:de_optimization_ns_faids}. The proposed hardware architectures, with both MS and NS-FAID decoding kernels,  are discussed in Section~\ref{sec:nsfaid_hw_architecture}. Numerical results are provided in Section~\ref{sec:nsfaid_implem_results}, and Section~\ref{ns_faids:sec:Conclusion} concludes the paper.



\section{Non-Surjective Finite Alphabet Iterative Decoders}
\label{sec:ns_faids}

\subsection{Preliminaries}

LDPC codes are defined by sparse bipartite graphs, comprising a set of variable-nodes (VNs), corresponding to coded bits, and a set of check-nodes (CNs), corresponding to parity-check equations. 
Finite Alphabet Iterative Decoders (FAIDs) are message-passing LDPC decoders that have been introduced in \cite{planjery2010iterative, planjery2011finite, planjery2013finite}. 
We state below the definition of a subclass of FAID decoders, which is less general than the one proposed in \cite{planjery2013finite}.
 Let $Q$ be a positive integer. A $(2Q+1)$-level FAID is a 4-tuple $\left({\cal M}, \Gamma, \Phi_v, \Phi_c\right)$, where:
\begin{itemize}
\item ${\cal M} = \{-Q,\dots,-1,0,+1,\dots,+Q\}$ is the alphabet of the exchanged messages, and is also referred to as the {\em decoder alphabet},
\item $\Gamma \subseteq {\cal M}$ is the input alphabet of the decoder, {\em i.e.}, the set of all possible values of the quantized soft information supplied to the decoder,
\item $\Phi_v$ and $\Phi_c$ denote the update rules for VNs and CNs, respectively.
\end{itemize}
We shall use $\m \in {\cal M}$ and $\gamma\in \Gamma$ to denote elements of ${\cal M}$ and $\Gamma$, respectively.
The CN-update function $\Phi_c$ is the same for any FAID decoder, and is equal to the update function used by the MS decoder. Precisely, for a CN of degree $d_c$, the update function $\Phi_c:  {\cal M}^{d_c-1} \rightarrow {\cal M}$ is given by:
\begin{equation}\label{eq:phi_c}
\hspace*{-2mm}\Phi_c\left(\m_1,\dots,\m_{d_c-1}\right) = \left(\prod_{i=1}^{d_c-1}\text{sgn}(\m_i) \right) \min_{i=1,\dots,d_c-1} |\m_i|\hspace*{-2mm}
\end{equation}

The VN-update function $\Phi_v:  \Gamma \times {\cal M}^{d_v-1} \rightarrow {\cal M}$, for a VN of degree $d_v$, is defined as:
\begin{equation}\label{eq:phi_v}
\Phi_v\left(\gamma, \m_1,\dots,\m_{d_v-1}\right) = F\left( \gamma + \sum_{j=1}^{d_v-1} \m_j  \right)
\end{equation}
where the function $F: \mathbb{Z} \rightarrow {\cal M}$ is defined based on a set of threshold values ${\cal T} = \{T_0, T_1,\dots,T_{Q+1}\} \subset \bar{\mathbb R}_+$, with $T_0 = 0$, $T_{Q+1} = +\infty$, and $T_i < T_j$ for any $i < j$:
\begin{equation}\label{eq:reframe_funct}
F(x) = \text{sgn}(x) \m, \ \ \text{where } \m \text{ is s.t. } T_\m \leq |x| < T_{\m+1}
\end{equation}

\noindent In Eq.~(\ref{eq:phi_v}), the variable $\gamma$ represents the {\em channel contribution}, {\em i.e.}, the quantized soft information that has been supplied to the decoder for the corresponding variable-node. 
The quantization method and its impact on FAIDs' decoding performance will be discussed in Section~\ref{sec:de_faid}.
It is also worth noting that in \cite{planjery2013finite}, FAIDs are introduced with a more general VN-update function $\Phi_v$, but the simpler definition (\ref{eq:phi_v}) that we use has many hardware implementation benefits, which will be described in later sections. Moreover, it can be easily seen that any non-decreasing odd function satisfies Eq.~(\ref{eq:reframe_funct}). Precisely, the following proposition holds. 
\begin{proposition}\label{prop:reframe_funct}
For any function $F:\mathbb{Z}\rightarrow {\cal M}$, there exists a threshold set ${\cal T}$ such that $F$ is given by Eq.~(\ref{eq:reframe_funct}), if and only if $F$ satisfies the following two properties:
\begin{itemize}
\item[(i)] $F$ is an odd function, {\em i.e.}, $F(-x) = -F(x)$, $\forall x\in \mathbb{Z}$
\item[(ii)] $F$ is non-decreasing, {\em i.e.}, $F(x) \leq F(y)$ for any $x < y$.
\end{itemize}
\end{proposition}
We note that the above proposition also implies that $F(0)=0$ and $F(x)\geq 0, \forall x > 0$. In this paper, we further extend the definition of FAIDs by allowing $F(0)$ to take on non-zero values. To ensure symmetry of the decoder, we shall write $F(0) = \pm \lambda$, with $\lambda \geq 0$, meaning that $F(0)$ takes on either $-\lambda$ or $+\lambda$ with equal probability. In the following, $F$ will be referred to as {\em framing function}.


As the focus of this work is on practical implementations, we will further assume that the sum  $\gamma + \sum_{j=1}^{d_v-1} \m_j$ in Eq.~(\ref{eq:phi_v}) is saturated to ${\cal M}$, prior to applying $F$ on it. Consequently, in the sequel we shall only consider framing functions  $F: {\cal M} \rightarrow {\cal M}$, and the VN-update function $\Phi_v$ from Eq.~(\ref{eq:phi_v}) is redefined as:
\begin{equation}\label{eq:phi_v_redef}
\Phi_v\left(\gamma, \m_1,\dots,\m_{d_v-1}\right) = F\left( s_{\cal M} \left(\gamma + \sum_{j=1}^{d_v-1} \m_j  \right)\right)
\end{equation}
where $s_{\cal M}:\mathbb{Z}\rightarrow {\cal M}$, $s_{\cal M}(x)= \mbox{sgn}(x)\min(|x|, Q)$, is the saturation function.
Since $F(-\m) = -F(\m), \forall \m\in {\cal M}$, $F$ is completely determined by the vector $[|F(0)|, F(1), ..., F(Q)]$, further referred to as the {\em Look-Up Table (LUT)} of $F$, which satisfies the following inequalities (Proposition~\ref{prop:reframe_funct}):
\begin{equation}\label{eq:framing_lut}
0 \leq |F(0)| \leq F(1) \leq \cdots \leq F(Q) \leq Q
\end{equation}

Summarizing, the subclass of FAIDs considered in this paper is defined by Eq.~(\ref{eq:phi_v_redef}), where $F$ is a framing function satisfying Eq.~(\ref{eq:framing_lut}). Furthermore, for any integer $q>0$, the expression {\em $q$-bit FAID} is used to refer to a $(2Q+1)$-level FAID, with $Q = 2^{q-1}-1$. It follows that messages exchanged within the FAID decoder are $q$-bit messages (including 1 bit for the sign). Finally, the message-passing iterative decoding process for a FAID with framing function $F$ is depicted in Algorithm~\ref{alg:faid}.


\begin{algorithm}[!t]
\caption{FAID decoding with framing function $F$}\label{alg:faid}
\begin{algorithmic}[0]
\State{Input:}  $\vect{y} = (y_1,\dots,y_N)$ 
\Comment{\small received word}
\State{Output:} $\hat{\vect{x}} = (\hat{x}_1,\dots,\hat{x}_N)$ \Comment{\small estimated codeword}  

\Init 
   \ForAll{$n=1,\dots,N$} 
          $\gamma_{n} = \displaystyle\text{quant}\left(\text{\sc llr}(x_n \mid y_n)\right)$;   
   \EndFor
   \ForAll{$m=1,\dots,M$ and $n\in{H}(m)$} 
          $\beta_{m,n} = 0$;
   \EndFor
\EndInit
\Iter 
   \ForAll{$n=1,\dots,N$ and $m\in{H}(n)$} \
       \Comment{{\bf\small VN-processing}}  \vspace*{1mm}
        \State $\alpha_{m,n} = \Phi_v\left(\gamma_n,  \beta_{m',n} \mid {m'}
		        \in{H}(n)\setminus m  \right)$;  \vspace*{1mm}
   \EndFor
   
   \ForAll{$m=1,\dots,M$ and $n\in{H}(m)$} \
       \Comment{{\bf\small CN-processing}}\vspace*{1mm}
       \State $\beta_{m,n}= \Phi_c\left(\alpha_{m,n'} \mid {n'}
	         \in{H}(m)\setminus n  \right)$; \vspace*{1mm}
   \EndFor

	\ForAll{$n=1,\dots,N$}\ 
	    \Comment{{\bf\small AP-update}}     \vspace*{1mm}   
        \State $\tilde{\gamma}_{n}=\displaystyle \gamma_n + \sum_{{m}
		        \in{H}(n)} \beta_{m,n}$;\vspace*{0mm}
	\EndFor
	
	\ForAll{$n=1,\dots,N$} $\hat{x}_n = \text{sign}(\tilde{\gamma}_{n})$; \ 
	    \Comment{\small hard decision}
	\EndFor
	
	\If{$\hat{\vect{x}}$ is a codeword} exit the iteration loop \Comment{\small syndrome check}
	\EndIf
\EndIter
\vspace*{-1mm}\State \hspace*{-3mm}\rule{\EndIterLen}{0.5pt}
{\footnotesize
\State Notation: ${H}$ -- bipartite graph of the LDPC code, with $N$ VNs and $M$ CNs; 
${H}(n)$ -- set of CNs connected to VN $n$; ${H}(m)$ -- set of VNs connected to CN $m$; $\alpha_{m,n}$ -- message from VN $n$ to CN $m$; $\beta_{m,n}$ -- message from CN $m$ to VN $n$; $\gamma_n$ and $\tilde{\gamma}_n$ -- a priori and a posteriori LLR of VN $n$; \text{quant} -- quantization map (see also Section~\ref{sec:de_faid}). 
%
%
}

\end{algorithmic}
\end{algorithm}


\smallskip \noindent 

\subsection{Non-Surjective FAIDs}

The finite alphabet MS decoder is a particular example of FAID, with framing function $F: {\cal M} \rightarrow {\cal M}$ being the identity function.  Using Eq.~(\ref{eq:framing_lut}) it can be easily verified that this is the only FAID for which the framing function $F$ is {\em surjective} (or equivalently {\em bijective}, since ${\cal M}$ is finite). For any other framing function $F$, there exists at least one element of ${\cal M}$, which is not in the image of $F$.  The class of FAIDs defined by non-surjective framing functions\footnote{Although we restrict the study in this paper to non-surjective framing functions $F: {\cal M} \rightarrow {\cal M}$, it can be readily extended to non-surjective framing functions $F: \mathbb{Z} \rightarrow {\cal M}$. While such an extension would widen the class of NS-FAIDs, our preliminary investigations have shown that the best NS-FAIDs are actually those within the subclass of NS-FAIDs defined by $F: {\cal M} \rightarrow {\cal M}$, investigated in this paper. } is investigated in this section. 

\begin{definition}
 The {\em weight} of a framing function $F: {\cal M} \rightarrow {\cal M}$, denoted by $W$, is the number of distinct entries in the vector $[|F(0)|, F(1), ..., F(Q)]$. It follows that $1\leq W\leq Q+1$.  By a slight abuse of terminology, we shall also refer to $W$ as the weight of the NS-FAID. We further define the {\em framing bit-length} as $w=\lceil\log_2(W)\rceil+1$.
\end{definition}

\begin{definition}
A {\em non-surjective FAID} (NS-FAID) is a FAID of weight $W < Q+1$. Hence the framing function $F$ is non-surjective, meaning that the image set of $F$ is a strict subset of ${\cal M}$. 
\end{definition} 

Table~\ref{tab:frame_funct_lut} provides two examples of $q=4$-bit NS-FAIDs (hence $Q=7$), both of which are of weight $W=4$, hence framing bit-length $w = \lceil\log_24\rceil+1 =3$. Note that $F_1$ maps $0$ to $0$, while $F_2$ maps $0$ to $\pm1$. The image sets of $F_1$ and $F_2$ are $\text{Im}(F_1) = \{0,\pm1,\pm3,\pm7\}$ and $\text{Im}(F_2) = \{\pm1,\pm3,\pm4,\pm7\}$.

\begin{table}[!t]
\centering
\caption{Examples of $4$-bit framing functions of weight $W=4$\vspace*{-2mm}}
\label{tab:frame_funct_lut}
\begin{tabular}{|c||c|c|c|c|c|c|c|c|}
\hline
$\m$      & $0$ & $1$ & $2$ & $3$ & $4$ & $5$ & $6$ & $7$ \\
\hline
$F_1(\m)$ & $0$ & $1$ & $1$ & $3$ & $3$ & $3$ & $7$ & $7$ \\
\hline
$F_2(\m)$ & $\pm1$ & $1$ & $1$ & $3$ & $3$ & $4$ & $4$ & $7$ \\
\hline
\end{tabular}
\end{table}  

The main motivation for the introduction of NS-FAIDs is that they allow reducing the size of the memory required to store the exchanged messages. Clearly, for a NS-FAID with framing bit-length $w$, the exchanged messages can be represented using only $w$ instead of $q$ bits (including $1$ bit for the sign). Moreover, as a consequence of the message size reduction, the size of the interconnect network that carries the messages from the memory to the processing units is also reduced.


\begin{proposition}\label{prop:nb_faids}
The number of $(2Q+1)$-level NS-FAIDs of weight $W$ is given by:
\begin{equation}
N_{\text{\rm NS-FAID}}(Q, W) = \binom{Q}{W-1}\binom{Q+1}{W}
\end{equation}
where $\binom{y}{x}$ denotes the binomial coefficient.
\end{proposition}

\subsection{Examples of NS-FAIDs}

As mentioned above, if the framing function $F$ is the identity function, then the corresponding FAID is equivalent to the MS decoder with finite alphabet ${\cal M}$. Some examples of NS-FAIDs are provided below.

\smallskip\noindent{\bf Example 1.} Let $F:{\cal M}\rightarrow{\cal M}$ be defined by: 
\begin{equation}
F(\m) = \text{sgn}(\m)\max(|\m|-\theta, 0)
\end{equation} 
where $\theta \in \{1,\dots,Q-1\}$. Then, the corresponding NS-FAID  is the Offset Min-Sum (OMS) decoder with offset factor $\theta$.

\smallskip\noindent{\bf Example 2.} Let $F:{\cal M}\rightarrow{\cal M}$ be defined by: 
\begin{equation}F(\m) = \left\{\begin{array}{cl}
\m, & \text{if } |\m| \text{ is even}\\
\text{sgn}(\m)(|\m|-1), & \text{if } |\m| \text{ is odd}
\end{array} \right.
\end{equation} 
Then, the corresponding NS-FAID  is the Partially OMS (POMS) decoder from \cite{nguyenly2015fpga}.

\smallskip
Moreover, it can be seen that the MS-based decoders proposed in \cite{oh2010min, chandrasetty2012area} and the dual-quantization domain decoder proposed in \cite{abu2014low} are particular realizations of NS-FAIDs. 

\smallskip While the main reason behind the NS-FAIDs definition consists in their ability to reduce memory and interconnect requirements, we can also argue that they may allow improving the error correction performance (with respect to MS). This is the case of both OMS and POMS decoders mentioned above. Given a target message bit-length $w$
({\em e.g.}, corresponding to some specific memory constraint),  one may try to find the framing function $F$ of corresponding weight $W$, which yields the best error correction performance. The optimization of the framing function can be done by using the DE technique, which will be discussed in Section \ref{sec:de_faid}.

\smallskip Since $F$ is a non-decreasing function, it can be shown that the framing function $F$ can alternatively be applied at the CN-processing step (instead of VN-processing), while resulting in an equivalent decoding algorithm.
Whether $F$ is applied at the VN-processing or the CN-processing step is rather a matter of implementation. 
When $F$ is applied at the VN-processing step, both VN- and CN-messages belong to a strict subset of the alphabet ${\cal M}$, namely  ${\cal M}' = \text{Im}(F) \subset {\cal M}$. When $F$ is applied at the CN-processing step, only CN-messages belong to ${\cal M}'$. However, it is worth noting that many hardware implementations of Quasi-Cyclic (QC) LDPC decoders rely on a layered architecture, which only requires storing the check-node messages \cite{boutillon2014hardware}.

\subsection{Irregular NS-FAIDs}

In case of irregular LDPC codes, {\em irregular NS-FAIDs} are NS-FAIDs using different framing functions $F_{d_v}$ for  VNs of different degrees $d_v$. Framing functions $F_{d_v}$ may have different weights  $W_{d_v}$. In this case, messages outgoing  from degree-$d_v$ VNs can be represented by using only $w_{d_v} = \lceil\log_2(W_{d_v})\rceil+1$ bits. However, the message size reduction does not necessarily apply to CN-messages, due to the fact that a CN may be connected to VNs of different degrees. Let  ${\cal M}'_{d_v} = \text{Im}(F_{d_v})$. Then, messages outgoing from a CN $c$ can be represented by using: 
\begin{equation}\label{eq:cn_mess_size}
\left\lceil\log_2\left(\left| \cup_{d_v\in{\cal D}_c} {\cal M}'_{d_v} \right|\right)\right\rceil \mbox{ bits,}
\end{equation}
where ${\cal D}_c$ is the set of degrees of VNs connected to $c$, and $|\cdot|$ is used to denote the number of elements of a set.

Alternatively, it is also possible to define CN-irregular NS-FAIDs in a similar manner. However, in this work we only deal with VN-irregular NS-FAIDs, since most of the practical irregular LDPC codes are irregular on VNs, while almost regular (or semi-regular) on CNs. In order to reduce the size of the CN-messages, in Section~\ref{subsec:opt_irreg_nsfaid} we will further impose certain conditions on the framing functions $F_{d_v}$, by requiring their images being included in one another.

\subsection{Density Evolution Analysis}
\label{sec:de_faid}

For the sake of simplicity, we only consider transmission over {\em binary-input} memoryless noisy channels. We assume that the channel input alphabet is ${\cal X} = \{-1, +1\}$, with the usual convention that $+1$ corresponds to the $0$-bit and $-1$ corresponds to the $1$-bit, and denote by ${\cal Y}$ the output alphabet of the channel. 
We denote by $x\in{\cal X}$ and $y \in {\cal Y}$ the transmitted and received symbols, respectively. 

We further consider a function $\varphi : {\cal Y} \rightarrow \Gamma$ that maps the output alphabet of the channel to the input alphabet of the decoder, and set $\gamma =\varphi(y)$. Hence, $\varphi$ encompasses both the computation of the soft (unquantized) log-likelihood ratio (LLR) value and its quantization. We shall refer to $\varphi$ as {\em quantization map} and to $\gamma$ as the {\em input LLR} of the decoder.

For transmission over the binary-input Additive White Gaussian Noise (AWGN) channel, we shall consider that the decoder's input information as well as the exchanged messages are quantized on the same number of bits; therefore $\Gamma = {\cal M}$ unless otherwise stated. In this case, $y = x + z$, where $z$ is the white Gaussian noise with variance $\sigma^2$, and the quantization map $\varphi : {\cal Y} \rightarrow {\cal M}$ is defined by:
\begin{equation}\label{eq:quant_map_varphi}
\varphi(y) = [\mu\cdot y]_{\cal M}
\end{equation}
where $\mu > 0$ is a constant referred to as {\em gain factor}, and $[x]_{\cal M}$ denotes the closest integer to $x$ that belongs to ${\cal M}$ (see also \cite{mheich2016code} and the gain factor quantizer defined therein).

The objective of the DE technique is to recursively compute the probability mass function (pmf) of the exchanged messages,  through  the  iterative decoding  process.  This  is  done under  the   assumption  that   exchanged messages are independent, which holds  in  the  asymptotic  limit  of the code length. In this case, the decoding performance converges to the cycle free case. 
DE equations for the NS-FAID decoder can be derived in a similar way as for the finite-alphabet MS decoder \cite{mheich2016code}. 
%
%
%
%
Similar to \cite{mheich2016code}, the DE is used to compute the asymptotic error probability, defined as:
\begin{equation}
p_e^{(+\infty)} = \lim_{\ell \rightarrow +\infty} p_e^{(\ell)}
\end{equation}
where $p_e^{(\ell)}$ is the bit error probability at iteration $\ell$. 

For a target bit error probability $\eta > 0$, the {\em $\eta$-threshold} is defined as the worst channel condition for which decoding error probability is less than $\eta$. Assuming the binary-input AWGN  channel model, the $\eta$-threshold corresponds to the maximum noise variance $\sigma ^2$ (or equivalently minimum SNR), such  that the asymptotic error probability is less than $\eta$:
\begin{equation}\label{eq:eta_threshold}
\sigma^2_{\text{\rm thres}} (\eta) = \sup \left\{{{\sigma ^2} \mid p_e^{\left( { + \infty } \right)} \leq \eta} \right\}
\end{equation}
In case that $\eta =0$, the $\eta$-threshold is simply referred to as DE threshold \cite{richardson2001capacity}. However, the asymptotic decoding performance of finite-precision MS-based decoders is known to exhibit an error floor phenomenon at high SNR \cite{mheich2016code}. This makes the $\eta$-threshold definition more appropriate in practical cases, when the target bit error rate can be fixed to a practical non-zero value.

%

Finally, it is worth noting that the $\sigma^2_{\text{\rm thres}} (\eta)$ value depends on:
(i) the irregularity of the LDPC code, parametrized as usual by the degree distribution polynomials $\lambda(x)$ and $\rho(x)$ \cite{richardson2001capacity}, (ii) the NS-FAID, {\em i.e.}, the size of the decoder alphabet and the framing function $F$,
(iii) the channel quantizer $\varphi$, or equivalently the gain factor $\mu$ used in Eq.~(\ref{eq:quant_map_varphi}). Therefore, assuming that the degree distribution polynomials $\lambda(x)$ and $\rho(x)$ and the size of the decoder alphabet are fixed, we use the DE technique to jointly optimize the framing function $F$ and channel quantizer. 


  
\section{Density Evolution Optimization of NS-FAIDs}
\label{sec:de_optimization_ns_faids}

Throughout this section, we consider $q=4$-bit NS-FAIDs (hence, $Q=7$). To illustrate the trade-off between hardware complexity and decoding performance, we consider the optimization of both regular and irregular NS-FAIDs. 

\subsection{Optimization of Regular NS-FAIDs}
\label{subsec:opt_reg_nsfaid}

\begin{table}[!t]
\caption{Best NS-FAIDs for $(3,6)$-regular LDPC codes\vspace*{-2mm}}
\label{tab:best_reg_nsfaid}
\fontsize{9pt}{9pt}\selectfont
\centering
\renewcommand{\arraystretch}{1.35}
\resizebox{\linewidth}{!}{%
\begin{tabular}{c|l|c|c}
  \multicolumn{2}{c|}{ } &  F  & SNR-thres (dB) \\
  \hline
 $w=4$ & MS     & $[0, 1, 2, 3, 4, 5, 6, 7]$ & $1.643$ ($\mu=5.6$) \\
 \hline
  $w=3$ & $F(0) = 0$    &  $\mathbf{[    0, 1, 1, 3, 3, 3, 7, 7 ]}$ & $\mathbf{1.409}$ ($\mathbf{\mu=3.8}$) \\
        & $F(0) = \pm1$ &  $[ \pm1, 1, 1, 3, 3, 4, 4, 7 ]$ & $1.412$ ($\mu=5.1$) \\
        & $F(0) = \pm2$ &  $[ \pm2, 2, 2, 3, 3, 3, 4, 7 ]$ & $1.712$ ($\mu=7.1$) \\
        & $F(0) = \pm3$ &  $[ \pm3, 3, 3, 3, 3, 4, 5, 7 ]$ & $2.227$ ($\mu=10.0$) \\
 \hline
 $w=2$ & $F(0) = 0$     & $[    0, 0, 0, 0, 0, 6, 6, 6 ]$ & $2.251$ ($\mu=8.6$)  \\
        & $F(0) = \pm1$ & $\mathbf{[\pm1, 1, 1, 1, 1, 6, 6, 6]}$ & $\mathbf{1.834}$ ($\mathbf{\mu=6.4}$)  \\
        & $F(0) = \pm2$ & $[ \pm2, 2, 2, 2, 2, 2, 2, 7 ]$ & $1.911$ ($\mu=8.3$)  \\
        & $F(0) = \pm3$ & $[ \pm3, 3, 3, 3, 3, 3, 3, 7 ]$ & $2.014$ ($\mu=9.4$)  \\
 \hline
\end{tabular}}
\end{table}

In this section, we consider the optimization of regular NS-FAIDs for $(d_v=3, d_c=6)$-regular LDPC codes. We consider $q=4$-bit NS-FAIDs, with framing bit-length parameter $w \in\{2, 3\}$. According to Proposition~\ref{prop:nb_faids}, the number of NS-FAIDs is given by  
$N_{\text{\rm NS-FAID}}(q=4, w=3) = N_{\text{\rm NS-FAID}}(Q=7, W=4) = \binom{7}{3}\binom{8}{4} = 2450$, and $N_{\text{\rm NS-FAID}}(q=4, w=2) = N_{\text{\rm NS-FAID}}(Q=7, W=2) = \binom{7}{1}\binom{8}{2} = 196$. 

\newlength{\cwa}\newlength{\cwb}\newlength{\cwc}\newlength{\cwd}\newlength{\cwe}\newlength{\cwf}
\settowidth{\cwa}{Decoder}
\settowidth{\cwb}{NS-FAID-444}
\settowidth{\cwc}{Framing functions applied to}
\settowidth{\cwd}{(\& gain factor $\mu$)}
\settowidth{\cwe}{($+\!/\!-$\,dB)}
\settowidth{\cwf}{Memory size reduction (\%)}
\def\xxx{\renewcommand{\arraystretch}{.8}}
\def\nsfaid{\text{NS-FAID-}}
\def\lut{\text{LUT}}
\def\mueq{\mu\!=\!}

\begin{table*}[!t]
\caption{Hardware Complexity vs. Decoding Performance Trade-Off for Optimized Irregular NS-FAIDs\vspace*{-2mm}}
\label{tradeoff}
\fontsize{9pt}{9pt}\selectfont
\centering
\renewcommand{\arraystretch}{1.35}
$\begin{array}{*{8}{@{}c@{}|}c@{}}
 \hline
 \multirow{3}{1.2\cwb}{\xxx\centering\begin{tabular}{@{}c@{}}NS-FAIDs\\Ensemble\end{tabular}} & 
 \multicolumn{3}{@{}c@{}|@{}}{\text{Framing functions applied to}} & 
 \multirow{3}{1.1\cwd}{\xxx\centering\begin{tabular}{@{}c@{}}SNR-thres (dB)\\(\& gain factor $\mu$)\\
                                              @BER\,$=10^{-6}$ \end{tabular}}          &
 \multirow{3}{1.1\cwe}{\xxx\centering\begin{tabular}{@{}c@{}}SNR\\gain/loss\\ ($+\!/\!-$\,dB)\end{tabular}}  &
 \multicolumn{3}{@{}c@{}}{\text{Memory size reduction (\%)}} \\
 \cline{2-4}\cline{7-9}
  & \multirow{2}{.35\cwc}{\centering$d_v=2$}&\multirow{2}{.35\cwc}{\centering$d_v=3$}&\multirow{2}{.35\cwc}{\centering$d_v=6$} & 
  &   & \multirow{2}{.35\cwf}{\xxx\centering\begin{tabular}{@{}c@{}}VN-\\mess.\end{tabular}} & \multicolumn{2}{@{}c@{}}{\text{CN-messages}} \\
  \cline{8-9}
 & & & & & & & \multirow{1}{.35\cwf}{\centering uncomp.} & \multirow{1}{.35\cwf}{\centering comp.}\\
\hline 
  \nsfaid444\text{ (MS) } & \lut0  & \lut0  & \lut0  & 1.374\ (\mueq3.2) & \ \ 0.000 &   0.00 &   0.00 &   0.00 \\
  \nsfaid432 & \lut0  & \lut1  & \lut6 & 1.188\ (\mueq3.0) & +0.186 & -27.63 &   0.00 &   0.00 \\ 
 \nsfaid433 & \lut0  & \lut3  & \lut2  & 1.015\ (\mueq2.8) & +0.359 & -17.76 &   0.00 &   0.00 \\ 
 \nsfaid332 & \lut4 & \lut3  & \lut6 & 1.273\ (\mueq2.6) & +0.101 & -34.87 & -25.00 & -13.04 \\ 
 \nsfaid333 & \lut4 & \lut4 & \lut3  & 1.110\ (\mueq2.4) & +0.264 & -25.00 & -25.00 & -13.04 \\ 
 \nsfaid222 & \lut7 & \lut5 & \lut5 & 2.299\ (\mueq2.3) & -0.925 & -50.00 & -50.00 & -26.09 \\ 
 \hline
\end{array}$
\end{table*}

All regular NS-FAIDs have been evaluated by using the DE technique from Section~\ref{sec:de_faid}. Table~\ref{tab:best_reg_nsfaid} summarizes the best NS-FAIDs according to $w$ and $F(0)$ values\footnote{NS-FAIDs with $|F(0)| > 3$ have worse DE thresholds, thus they have not been included in the table.}, for $0\leq |F(0)| \leq 3$; DE thresholds ($\eta = 0$) and corresponding gain factors ($\mu$) are also reported. Best NS-FAIDs for $w=2$ and $w=3$ are emphasized in bold. For comparison purposes, the DE threshold of the $q = 4$-bit MS decoder is also reported: MS threshold is equal to $1.643$\,dB, for  $\mu=5.6$. For $w = 3$, it can be observed that best NS-FAIDs with $F(0) = 0$ or $F(0)=\pm{1}$ have better DE thresholds than the $4$-bit MS decoder. The best NS-FAID is given by the framing function $F = [0, 1, 1, 3, 3, 3, 7, 7]$ and its DE threshold is equal to $1.409$\,dB ($\mu=3.8$), representing a gain of $0.23$\,dB compared to $4$-bit MS.  For $w = 2$, the best NS-FAID is given by the framing function $F = [\pm{1}, 1, 1, 1, 1, 6, 6, 6]$ and its DE threshold is equal to $1.834$\,dB ($\mu=6.4$), which represents a performance loss of only $0.19$\,dB compared to $4$-bit MS. To emphasize the benefits of the proposed NS-FAIDs extension, we note that that for $w = 2$, best NS-FAIDs with $F(0)=\pm{1}$, $F(0)=\pm{2}$ or $F(0)=\pm{3}$ have better DE thresholds than the best NS-FAIDs with $F(0)=0$. The latter is given by the framing function $F = [0, 0, 0, 0, 0, 6, 6, 6]$ and its DE threshold is equal to $2.251$\,dB ($\mu=8.6$), thus resulting in a performance loss of approximately $0.61$\,dB compared to $4$-bit MS.

\subsection{Optimization of Irregular NS-FAIDs}
\label{subsec:opt_irreg_nsfaid}
As a case study, we consider the optimization of irregular NS-FAIDs for the WiMAX irregular LDPC codes with rate $1/2$ \cite{wimax} (of course, the proposed method can be applied to any other irregular codes in the same manner). The {\em edge-perspective} degree distribution polynomials are given by $\lambda \left( x \right) = 0.2895x + 0.3158{x^2} + 0.3947{x^5}$ and $\rho \left( x \right) = 0.6316{x^5} + 0.3684{x^6}$. Hence, VNs are of degree $d_v\in\{2,3,6\}$. For each VN-degree $d_v$, we consider that the corresponding framing function $F_{d_v}$ may be of any weight $W_{d_v}\in\{2, 4, 8\}$, corresponding to a framing bit-length $w_{d_v}\in\{2, 3, 4\}$. Hence, the total number of framing functions  is given by $N_{\text{\rm NS-FAID}}(7, 2)+ N_{\text{\rm NS-FAID}}(7, 4) + N_{\text{\rm NS-FAID}}(7, 8)= 2645$ (see Proposition~\ref{prop:nb_faids}). Since a different framing function may be applied for each VN-degree, it follows that the total number of irregular NS-FAIDs is equal to $2645^3 = 18\,504\,486\,125$. Clearly, even though we rely on DE, it is practically impossible to evaluate the decoding performance of all the irregular NS-FAIDs. To overcome this problem, we proceed as described below. 

\subsubsection{Optimization procedure}
First, we evaluate the DE thresholds of NS-FAIDs applying one and the same framing function  to all the variable-nodes, irrespective of their degree, which for simplicity will be referred to as {\em uniform} NS-FAIDs throughout this section. Uniform NS-FAIDs with framing bit-length $w=\{2, 3, 4\}$ are then sorted with increasing DE threshold value, from the best to the worst decoder. Note that the case $w=4$ represents a slight abuse of terminology, since there is only one such decoder, corresponding to the original MS decoder. We further denote by ${\cal U}^{\text(\rm best)}$-NS-FAID-$w$ the set of best uniform NS-FAIDs with framing bit-length $w$, determined as follows:
\begin{itemize}
\item ${\cal U}^{\text(\rm best)}$-NS-FAID-$2$ is comprised of the uniform NS-FAIDs with $w=2$, whose DE threshold is less than or equal to $5$\,dB; this represents $121$ decoders out of the total of $N_{\text{\rm NS-FAID}}(Q=7, w=2)=196$ decoders.
\item ${\cal U}^{\text(\rm best)}$-NS-FAID-$3$ is comprised of the uniform NS-FAIDs with $w=3$, whose DE threshold is less than or equal to $3$\,dB; this represents $946$ decoders out of the total of $N_{\text{\rm NS-FAID}}(Q=7, w=3)=2450$ decoders.
\item ${\cal U}^{\text(\rm best)}$-NS-FAID-$4$ is comprised of the MS decoder only; its DE threshold is equal to $1.374$\,dB.
\end{itemize} 

\noindent The limiting values of the DE thresholds for ${\cal U}^{\text(\rm best)}$-NS-FAID-$2$ and ${\cal U}^{\text(\rm best)}$-NS-FAID-$3$ are chosen such that the number of selected NS-FAIDs in each set is small enough. 

For irregular NS-FAIDs, we denote by NS-FAID-$w_2w_3w_6$ the ensemble of NS-FAIDs defined by a triplet of framing functions $(F_2, F_3, F_6)$, corresponding to variable node degrees $d_v\!\!=\!\!2,3,6$, with framing bit-lengths $w_2, w_3, w_6$. Since $w_2, w_3, w_6 \in\{2,3,4\}$, there are $27$ such ensembles. Since the number of NS-FAIDs in these ensembles can be very large, we only evaluate part of them, by further imposing the following two constraints: 

\noindent \textit{Decoding performance constraint}: We only consider irregular NS-FAIDs defined by triplets of  framing functions $(F_2, F_3, F_6)$, such that $F_{d_v}\in {\cal U}^{\text(\rm best)}\text{-NS-FAID-}w_{d_v}$, for any $d_v\in\{2, 3, 6\}$.

\noindent \textit{Memory size reduction constraint}: We further impose the following inclusion constraint between the image sets of framing functions used for different VN-degrees. Let $w^{\text{(max)}} = \max_{d_v}w_{d_v}$ and $d_v^{\text{(max)}} = \mathop{\rm argmax}_{d_v}w_{d_v}$. We impose that ${\mathop{\rm Im}\nolimits} \left( {{F_{d_v}}} \right) \subseteq {\mathop{\rm Im}\nolimits} \left( {{F_{d_v^{\text{(max)}}}}} \right)$, $\forall d_v\in\{2, 3, 6\}$. According to Eq.~(\ref{eq:cn_mess_size}),  this constraint ensures that CN-messages can be represented by using only $w^{\text{(max)}}$ bits, which is particularly suitable for layered architectures.

The number of irregular NS-FAIDs that satisfy the above two constraints 
is equal to $N_{\text{NS-FAIDs}}^{\text{irregular}} = 7\,017\,762$.

\def\lut{\text{\footnotesize LUT}}
\begin{table}[!t]
\caption{LUTs used by NS-FAIDs in Table \ref{tradeoff}\vspace*{-2mm}}
\label{tab:luts}
\fontsize{9pt}{9pt}\selectfont
\centering
$\begin{array}{@{\ }c@{\ }|*{8}{@{\;}c@{\;}}}
 \m & \lut0  & \lut1  & \lut2  & \lut3  & \lut4 & \lut5 & \lut6 & \lut7 \\ 
 \hline
 0 &   0  &   0  &   0  &   0  &   0  & \pm1 & \pm1 & \pm1 \\ 
 1 &   1  &   0  &   1  &   1  &   1  &   1  &   1  &   1  \\ 
 2 &   2  &   2  &   1  &   1  &   1  &   1  &   1  &   1  \\ 
 3 &   3  &   2  &   2  &   3  &   3  &   1  &   1  &   5  \\ 
 4 &   4  &   3  &   2  &   3  &   3  &   5  &   7  &   5  \\ 
 5 &   5  &   3  &   7  &   3  &   7  &   5  &   7  &   5  \\ 
 6 &   6  &   7  &   7  &   7  &   7  &   5  &   7  &   5  \\ 
 7 &   7  &   7  &   7  &   7  &   7  &   5  &   7  &   5  \\ 
 \hline\hline
 w &   4  &   3  &   3  &   3  &   3  &   2  &   2  &   2  \\ 
 \hline
\end{array}$
\end{table}

\subsubsection{Density Evolution evaluation}
For each of the $N_{\text{NS-FAIDs}}^{\text{irregular}}$ irregular NS-FAIDs, we compute its decoding threshold for a target bit error rate $\eta=10^{-6}$, using the DE technique from Section~\ref{sec:de_faid}. The threshold computation also encompasses the optimization of the channel gain factor $\mu$. Hence, for each NS-FAID, we first determine the gain factor $\mu$ that maximizes the $\eta$-threshold defined in Eq.~(\ref{eq:eta_threshold}). The corresponding $\eta$-threshold value is then reported as the $\eta$-threshold of the NS-FAID.

Density evolution results for the MS decoder (indicated as NS-FAID-444), as well as for the NS-FAIDs with the best $\eta$-thresholds from five NS-FAID-$w_2w_3w_6$ ensembles, are shown in Table~\ref{tradeoff}: the framing functions used for VN-degrees $d_v=2,3,6$ are shown in columns 2, 3, and 4, while the $\eta$-threshold value (in dB) and the corresponding gain factor $\mu$ are shown in column 5. The SNR gain\,($+$) or loss\,($-$) reported in column 6 corresponds to the differences between the SNR threshold of the MS decoder (NS-FAID-$444$) and the SNR threshold of the best NS-FAID-$w_2w_3w_6$. The memory size reduction of the NS-FAID-$w_2w_3w_6$ decoders compared to the MS decoder is reported in columns 7-9, for both VN and CN messages. For CN messages, two possibilities are considered, according to whether they are stored in an {\em uncompressed} or {\em compressed} format, where  compressed format means that only the signs, first minimum, second minimum, and index of the first minimum are stored \cite{wang2007memory}. Finally, the framing functions' LUTs of the best NS-FAID-$w_2w_3w_6$ decoders are reported in Table~\ref{tab:luts}. 

\begin{figure}[!t]
\centering
\includegraphics [width=\linewidth] {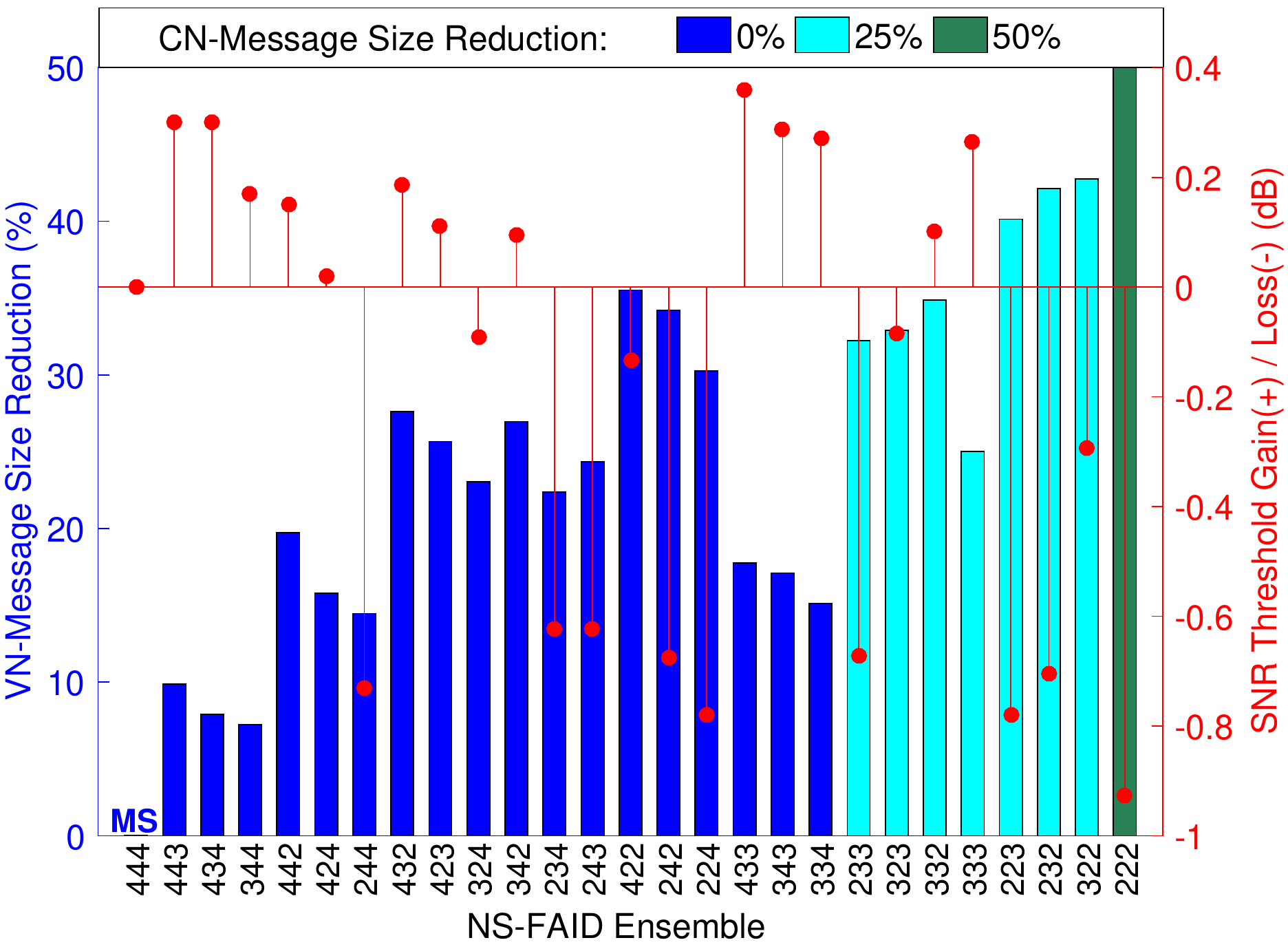}
\caption{Memory size reduction vs. decoding performance}
\label{fig_tradeoff}
\end{figure}

Fig.~\ref{fig_tradeoff} captures the trade-off between decoding performance and memory size reduction, for each of the 27 NS-FAID-$w_2w_3w_6$ ensembles. For each ensemble, we select the NS-FAID with the best threshold, and indicate the corresponding memory gains and decoding performance.
The height of vertical bars indicates the VN memory size reduction (values on the left vertical axis), while their color indicates the CN memory size reduction (uncompressed CN message storage is assumed in the legend). The red stems indicate the SNR threshold gain or loss compared to MS decoder (values on the right vertical axis). It can be seen that the NS-FAID-332 decoder allows a significant memory size reduction for both variable- and check-node messages, while still performing $0.1$\,dB ahead the MS decoder. The NS-FAID-433 decoder is also a very good candidate for applications requiring increased decoding performance: it achieves the best SNR gain ($0.36$\,dB), while providing a VN memory size reduction by $17.76\%$ with respect to the MS decoder.

Finally, it is worth noticing that the reported memory size reductions do not necessarily translate as such in hardware implementations, for several reasons. First, depending on the hardware architecture, VN messages may or may not be stored in a dedicated memory. For instance, layered architectures only require the storage of CN messages, which can be further stored in either a compressed or uncompressed format. Moreover, VN processing units (VNUs) need to be equipped with a framing and a deframing module, which may offset part of the promised gains. This is even more true in case of irregular codes, for which some VNUs may need to implement more than one single framing function, since the same VNU may be reused to process VN of different degrees (except for fully parallel architectures). To assess the gains of NS-FAIDs in practical implementations, the integration of the NS-FAID mechanism (framing/deframing modules) into two different  MS decoder architectures is discussed in the next section, and ASIC synthesis results for the CMOS $65$\,nm process technology are provided in Section~\ref{sec:nsfaid_implem_results}. 

\section{Hardware Architectures for NS-FAID Decoders}
\label{sec:nsfaid_hw_architecture}
In this section we propose two layered decoder architectures for QC-LDPC codes, with both MS and NS-FAIDs decoding kernels. Proposed architectures target high-throughput, while ensuring an efficient use of the hardware resources. Two possible approaches to achieve high-throughput are explored, consisting in either pipelining the datapath, or increasing the hardware parallelism. 


We consider a QC-LDPC code defined by a base matrix $B$ of size $R \times C$, and expansion factor $z$, corresponding to a parity check matrix $H$ of size $M\times N$, with $M=zR$ and $N=zC$. With the notation from Section~\ref{sec:de_faid}, we denote by  $\vect{x} = (x_1,\dots,x_N) \in{\cal X}^N$ the transmitted codeword,  by $\vect{y} = (y_1,\dots,y_N)\in {\cal Y}^N$ the received word, and by $(\gamma_1,\dots,\gamma_N)$ the input LLRs of the decoder, where $\gamma_n =\varphi(y_n)$ and $\varphi$ is the quantization map. VN and CN messages are denoted by $\alpha_{m,n}$ and $\beta_{m,n}$, respectively, and the {\em A Posteriori (AP)-LLR} of a VN $n$ is denoted by ${\tilde \gamma}_n$. Hence, $\tilde{\gamma}_n = \gamma_n + \sum_{m\in H(n)} \beta_{m,n}$, where the sum is taken over all the check-nodes $m$ connected to $n$, denoted by $m\in H(n)$. Input LLRs and exchanged messages are quantized on $q$-bits, while AP-LLRs are assumed to be quantized on $\tilde{q}$ bits, with $\tilde{q} > q$.

A {\em decoding layer} (or simply referred to as a layer) consists of one or several consecutive rows of $B$, assuming that they do not overlap, {\em i.e.}, each column of $B$ has at most one non-negative entry within each layer. It is assumed that the same number of rows of $B$ participate in each decoding layer, which is denoted by $\rpl$ (rows per layer). Hence, the number of decoding layers is given by $L = R/\rpl$. We further define $Z = z\times \rpl$, corresponding to the number of rows of $H$ (parity checks) within one decoding layer, and referred to as the parallelism degree of the hardware architecture. 
To ease the description of the hardware architectures proposed in this section, we shall assume that all CNs have the same degree, denoted by $d_c$. However, no assumptions are made concerning VN degrees. 
We present each architecture assuming the MS decoding kernel is being implemented, then we discuss the required changes in order to integrate the NS-FAID decoding kernel.  

\subsection{Pipelined architecture}
\label{subsec:pipelined_arch}


\begin{figure*}[!t]
\centering
\includegraphics[width=0.85\linewidth]{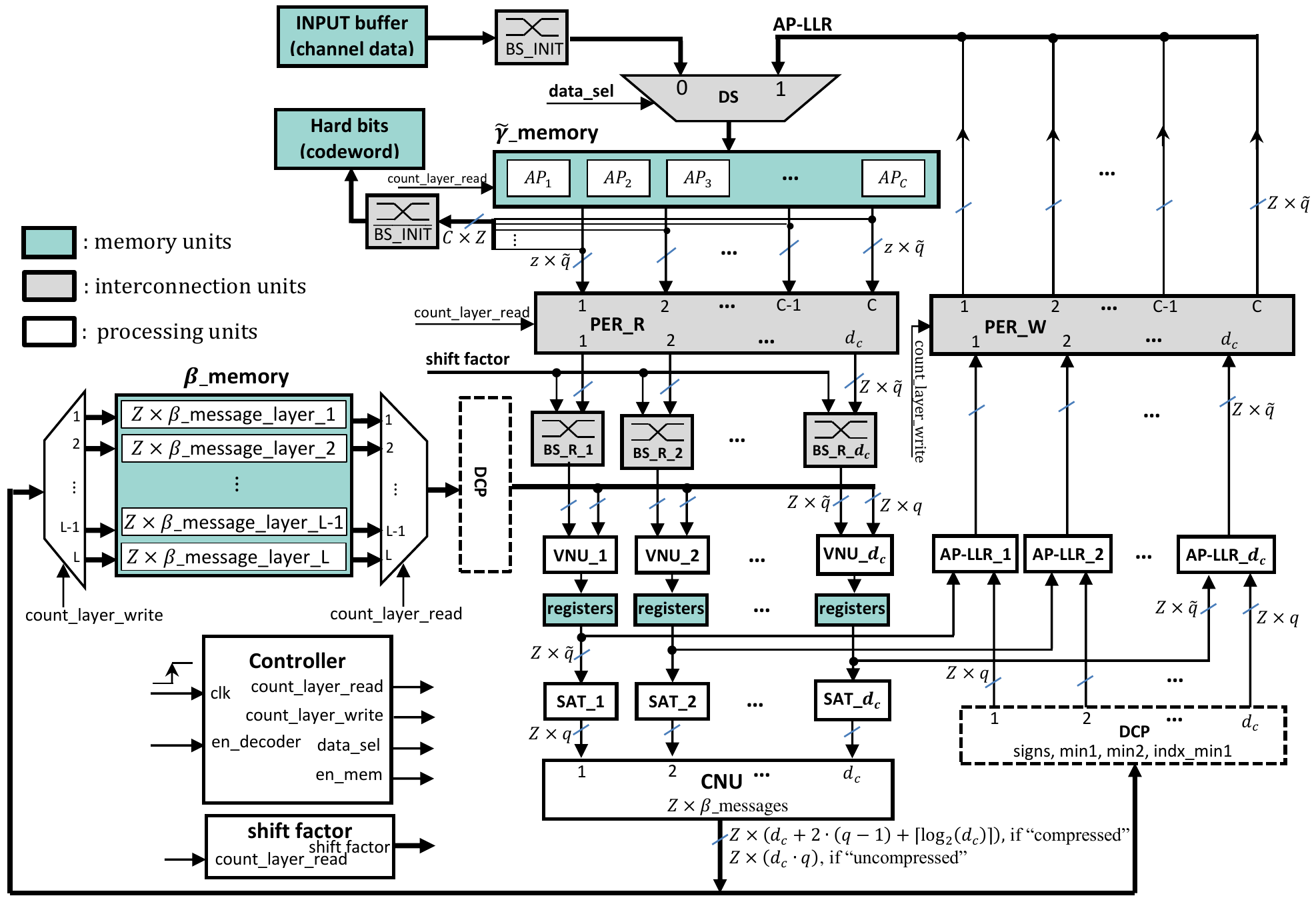}
\caption{Block Diagram of the Proposed Pipelined Architecture with MS-kernel}
\label{subfig:hw_architecture_2cc_long}
\end{figure*}

The proposed architecture with MS decoding kernel is detailed in Fig.~\ref{subfig:hw_architecture_2cc_long}. A high-level representation is also shown in Fig.~\ref{subfig:hw_architecture_2cc_short}, for both MS and NS-FAID decoding kernels. 
The architecture is optimized so that to reduce the critical path. In particular, we completely reorganize the interconnect network (barrel shifters BS\_INIT and BS\_R, see below), thus removing the need for a barrel shifter on the writing data back side. The main blocks of the architecture are discussed below.


\subsubsection*{Input/Output buffers} The input buffer, implemented as a number of Serial Input Parallel Output (SIPO) shift registers, is used to store the input LLR values ($\gamma_n$) received by the decoder. The output buffer, is used to store the hard bit estimates of the decoded word. Input/output buffers allow data load/offload  operations to take place concomitantly with the decoding of the current codeword. 

\subsubsection*{Memory blocks} Two memory blocks are used, one for AP-LLR values ($\tilde{\gamma}$\_memory) and one for CN-messages ($\beta\_$memory).   
$\tilde{\gamma}$\_memory is implemented by registers, in order to allow massively parallel read or write operations. It is organized in $C$ blocks, denoted by $AP_j\,(j=1,\cdots,C)$, corresponding to the columns of the base matrix, each one consisting of $z \times \tilde{q}$ bits. Data are read from/written to blocks corresponding to non-negative entries in the decoding layer being processed. 
 $\beta\_$memory is implemented as a dual port Random Access Memory (RAM), in order to support pipelining, as explained below. Each memory word consists of $Z$ $\times$ $\beta\_$messages, corresponding to one decoding layer. Depending on the Check Node Unit (CNU) implementation, $\beta\_$messages can be either ``uncompressed'' ({\em i.e.}, for a check-node $m$, the corresponding $\beta$\_message is given by the $d_c$ values $[\beta_{m,n_1}, \dots, \beta_{m,n_{d_c}}]$, where $n_1, \dots, n_{d_c}$ denote the variable nodes  connected to $m$) or ``compressed''  ({\em i.e.}, for a check-node $m$, the corresponding $\beta$\_message is given by the signs of the above $\beta_{m,n_i}$ messages, their first and second minimum, denoted by $\msminone$ and $\msmintwo$, and the index of the first minimum, denoted by $\msidxone$) \cite{wang2007memory}.

%

\subsubsection*{Read and Write Permutations (PER\_R, PER\_W)} PER\_R permutation is used to rearrange the data read from $\tilde{\gamma}$\_memory, according to the processed layer, so as to ensure processing by the proper VNU/CNU. PER\_W block operates oppositely to PER\_R.

\subsubsection*{Barrel Shifters (BS\_INIT, BS\_R)} Barrel shifters are used to implement the cyclic (shift) permutations, according to the non-negative entries of the base matrix. 
The $\tilde{\gamma}$\_memory is initialized from the input LLR values stored in the input buffer. However, input LLR values are shifted by BS\_INIT block before being written to the $\tilde{\gamma}$\_memory, according to the {\em last} non-negative shift factor on the corresponding base matrix column. BS\_R blocks are then used to shift the LLR values read from the $\tilde{\gamma}$\_memory, such that to properly align them with the appropriate VNU. 
Note that there are $d_c$ BS\_R\_$\{1,\dots,d_c\}$ blocks. In case $\rpl=1$, a decoding layer corresponds to a row of $B$, and each BS\_R block is used to shift the LLR values within one of the $d_c$ columns with non-negative entries in the current row. Let $i$ be the index of the current row. The cyclic shift implemented by a BS\_R block, corresponding to a column $j$ with $b_{i,j}\geq 0$, is given by  $-b_{i',j}+b_{i,j}$, where $b_{i',j}$ is the  previous  non-negative entry in column $j$ ({\em i.e.}, the previous row $i'$ with $b_{i',j}\geq 0$). In case $\rpl > 1$, each BS\_R block actually consists of $\rpl$ sub-blocks as above, with one sub-block for each row in the layer. The values of the cyclic shifts are computed offline for each layer $\ell$. This eliminates the need for data write-back barrel shifters, thus reducing the critical path of the design. Finally, the $\overline{\text {BS\_INIT}}$ block operates oppositely to BS\_INIT, and is used to shift back the hard decision bits into appropriate positions.


\subsubsection*{Variable Node Units (VNUs) and AP-LLR Units} These units compute VN-messages ($\alpha_{m,n}$) and AP-LLR values ($\tilde{\gamma}_n$). Each VN message is computed by subtracting the corresponding CN message from the AP-LLR value, that is $\alpha_{m,n} = \tilde{\gamma}_n - \beta_{m,n}$. This operation is implemented by a $\tilde{q}$-bit subtractor, hence the $\alpha_{m,n}$ value outputted by the VNU is quantized on $\tilde{q}$ bits. The AP-LLR value is updated by the AP-LLR unit, by $\tilde{\gamma}_n = \alpha_{m,n} + \beta_{m,n}^{\text{new}}$, where
$\beta_{m,n}^{\text{new}}$ is the corresponding CN message computed at the current iteration (see below).

\subsubsection*{Saturators (SATs)} Prior to CNU processing, $\alpha_{m,n}$ values are saturated to $q$ bits.

\subsubsection*{Check Node Units (CNUs)} These processing units compute the CN-messages ($\beta_{m,n}$). For simplicity, Fig. \ref{subfig:hw_architecture_2cc_long} shows one CNU block with $d_c$ inputs, each one of size $Z \times q$ bits. Thus, this block actually includes $Z$ computing units, used to process in parallel the $Z$ check-nodes within one layer. The CNU is implemented by using either: (i) the high-speed low-cost tree-structure (TS) approach proposed in \cite{wey2008algorithms} for ``compressed'' CN-messages, or (ii) comparator trees for ``uncompressed'' CN-messages.

\subsubsection*{Decompress (DCP)} This block is only used in case that the CN-messages are in compressed format (signs, min1, min2, indx\_min1). It converts the $\beta$\_messages from compressed to the uncompressed format. 

\subsubsection*{Controller} This block generates control signals such as {\em count\_layer\_read}, {\em count\_layer\_write} to indicate which layers are being processed, {\em write\_en} to enable data writing, etc. It also controls the synchronous execution of the other blocks.

\subsubsection*{Pipelining} 
\begin{figure}[!t]
\centering
\includegraphics [width=\linewidth]{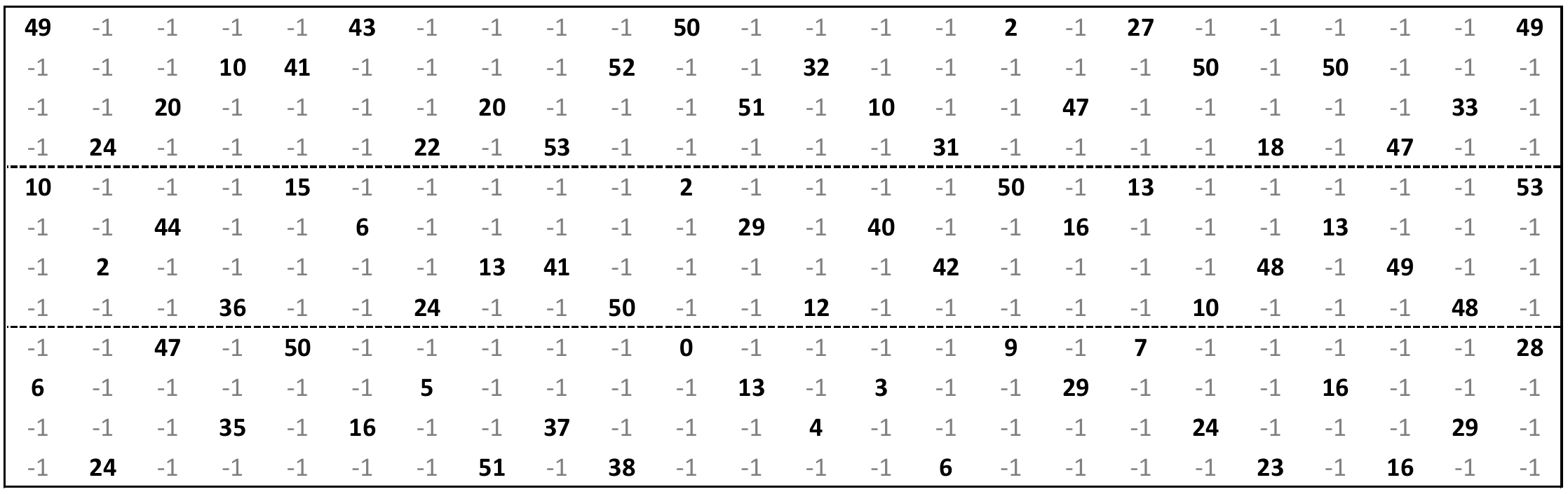}
\caption{Base matrix of the $(3,6)$-regular QC-LDPC code}
\label{fig:base_matrix_regular}
\end{figure}

To increase the operating frequency, the data path is pipelined by adding a set of registers after the VNU-blocks. The timing schedule is shown in Fig.~\ref{subfig:hw_architecture_2cc_short}, where the two pipeline stages (P1 and P2) are indicated by purple and brown arrows. Hence, processing one layer takes 2 clock cycles, but at each clock cycle the two pipeline stages work on two consecutive layers of the base matrix. This imposes specific constraints on the base matrix, as consecutive layers must not overlap, in order to avoid $\tilde{\gamma}$\_memory conflicts (note that memory stall cycles would cancel the pipelining effect). An example of $d_c=6$ regular base matrix without overlap between consecutive layers  is given in Fig.~\ref{fig:base_matrix_regular}, assuming that each layer corresponds to one row of the base matrix.


\subsubsection*{\bf \em Regular NS-FAID decoding kernel} 
\begin{figure*}[!t]
\centering
\subfigure[Pipelined Architecture]{\includegraphics[scale=0.55]{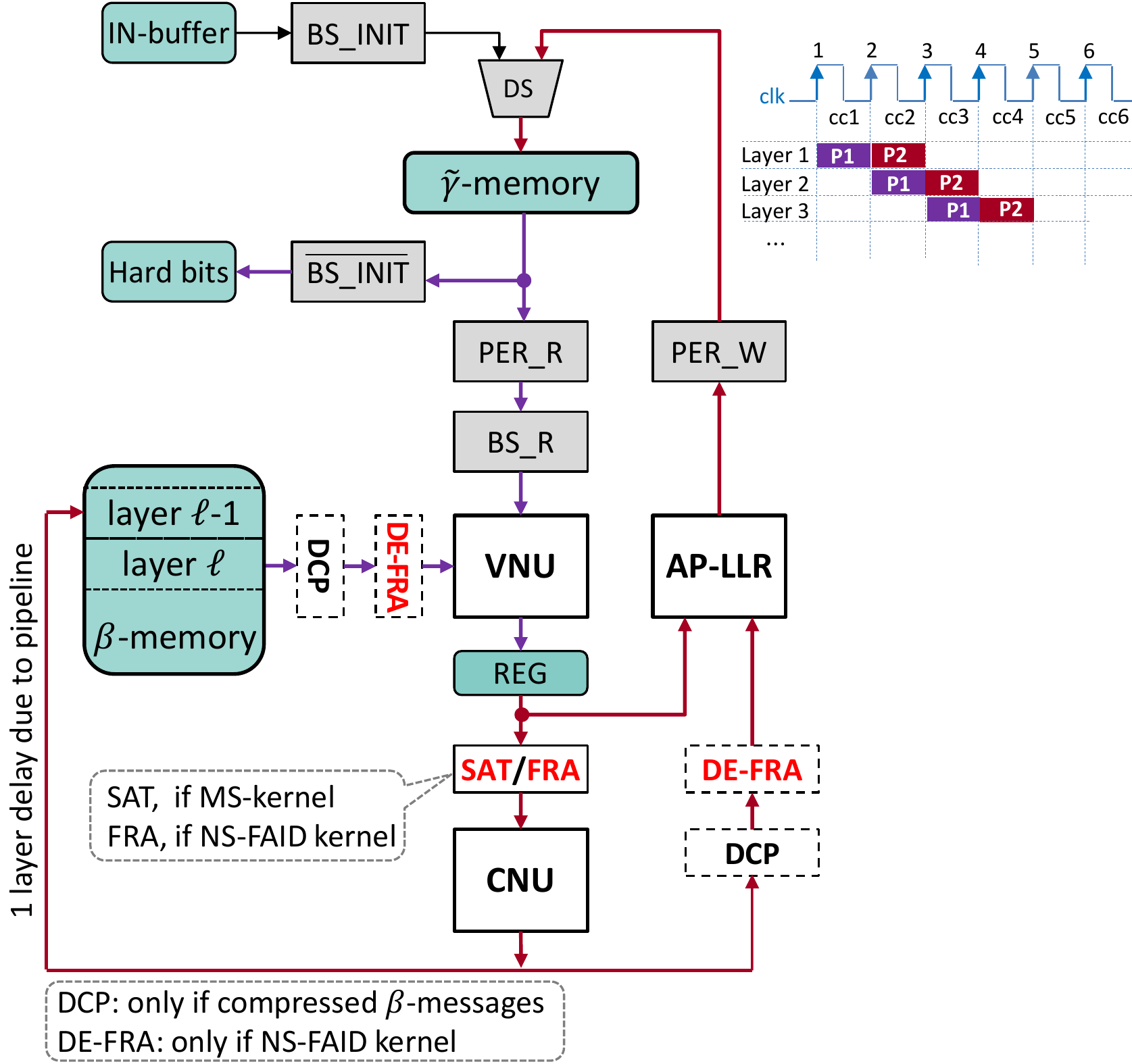}\label{subfig:hw_architecture_2cc_short}}\hspace*{-10mm}\rule[-5mm]{0.1mm}{65mm}\rule[59.9mm]{12mm}{0.1mm}\rule[59.8mm]{0.1mm}{26mm}\raisebox{63mm}{\hspace*{-25mm}{\footnotesize (timing schedule)}}\hfill%
\subfigure[Full-Layer Architecture]{\includegraphics[scale=0.55]{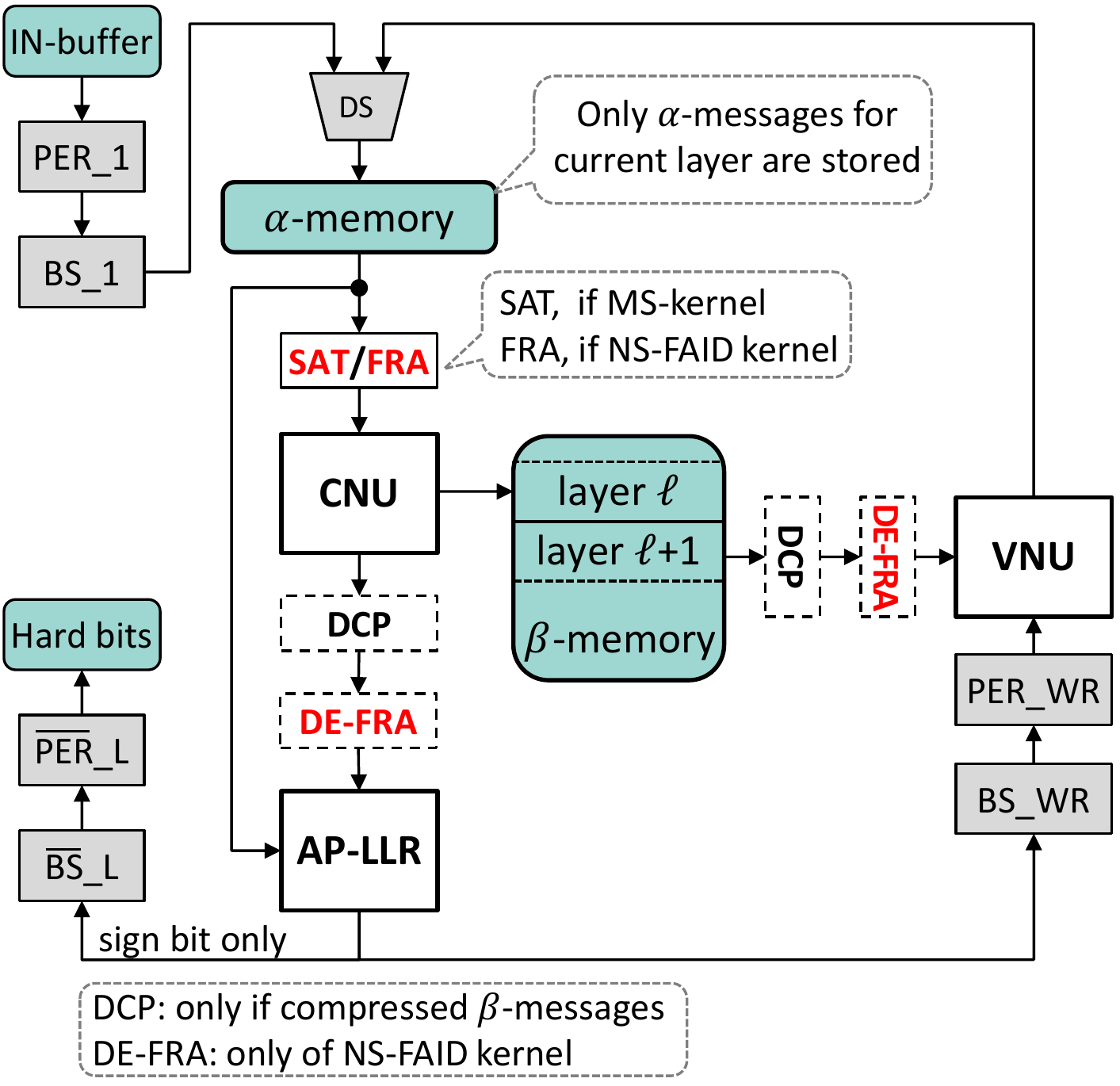}\label{fig_MS_NS_FAID_1cc_short}}
\caption{High-Level Description of the Proposed HW Architectures, with both MS and NS-FAID kernels}
\label{fig:hw_architecture_high_level}
\end{figure*}

The changes required to integrate a regular NS-FAID decoding kernel, with framing function $F$, are shown in Fig.~\ref{subfig:hw_architecture_2cc_short}. First, the Saturation (SAT) block used within the MS-decoding kernel is replaced by a Framing (FRA) block.  Note that the output of the VNU consists of $\tilde{q}$-bit (unsaturated) VN-messages. Hence, the FRA block actually implements the concatenation of the following operations, corresponding to $F\circ s_{\cal M}$ in Eq.~(\ref{eq:phi_v_redef}): 
\begin{equation}
\resizebox{\linewidth}{!}{$[-\tilde{Q},\dots,+\tilde{Q}] \stackrel{s_{\cal M}}{\longrightarrow} [-{Q},\dots,+{Q}] \stackrel{F}{\longrightarrow} \text{Im}(F) \stackrel{\sim}{\longrightarrow} [-{W},\dots,+{W}]$,}
\end{equation}
where $[-\tilde{Q},\dots,+\tilde{Q}]$ is the alphabet of unsaturated messages ($\tilde{Q} = 2^{\tilde{q}-1}-1$), $F$ is the framing function being used, $\text{Im}(F)$ is the image of $F$ (which is a subset of $[-{Q},\dots,+{Q}]$ according to the framing function definition), and the last operation consists of a re-quantization of the $\text{Im}(F)$ values on a number of $w$-bits, where $w=\lceil\log_2(W)\rceil+1$ is the framing bit-length. The De-framing (DE-FRA) block simply converts back from $w$-bit to $q$-bit values ($[-{W},\dots,+{W}]\stackrel{\sim}{\rightarrow}\text{Im}(F)\subset[-{Q},\dots,+{Q}] $), {\em i.e.}, it inverts the re-quantization operation above. Although we have to add the de-framing blocks, the reduction of the CN-messages size may still save significant hardware resources, as compared to MS decoding. This will be discussed in more details in Section~\ref{sec:nsfaid_implem_results}.

\subsubsection*{\bf \em Irregular NS-FAID decoding kernel} First, we note that the pipeline architecture proposed in the section can be applied to the WiMAX QC-LDPC code with rate $1/2$, considered in Section~\ref{subsec:opt_irreg_nsfaid}, by assuming that each decoding layer consists of one row of the base matrix. Indeed, it is known that for this code, the rows of the base matrix can be reordered, such that any two consecutive rows do not overlap \cite{zhang2015full}. 

Regarding the integration of an irregular NS-FAID decoding kernel, the same framing (FRA) or de-framing (DE-FRA) block is reused for several VNs, which may be of different degrees. This may require several framing functions to be implemented within the FRA/DE-FRA blocks, thus increasing the hardware complexity. To overcome this problem, one may change the way the VNs are mapped to the processing units, by reordering the columns of the base matrix processed within each decoding layer. We determine offline such a reordering for each decoding layer, so as to minimize the number of FRA/DE-FRA blocks implementing more than one single framing function. Hence, the PER\_R and PER\_W blocks, which ensure the proper alignment between data and processing units, are redefined accordingly. 

\begin{figure}[!t]
\centering
\includegraphics[width=.85\linewidth]{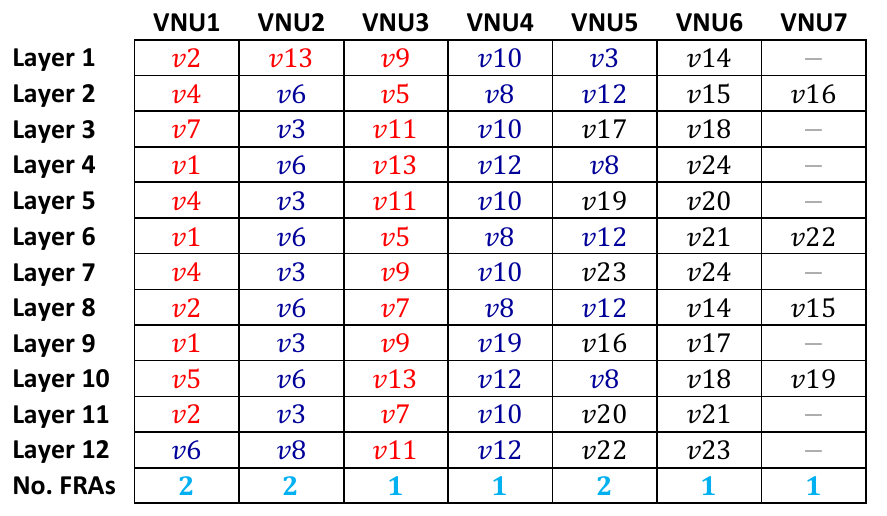}
\caption{Mapping between VNs and VNUs. In black: VNs of degree $2$, in red: VNs of degree $3$, in blue: VNs of degree $6$.}
\label{fig:column_reorder}
\end{figure}

The optimal mapping between VNs and VNUs is shown in Fig.~\ref{fig:column_reorder}, for the base matrix with reordered rows from \cite{zhang2015full}. For each VNU, we indicate the index of the VN (or equivalently, base matrix column) processed by the VNU, within each decoding layer. The last row of the table indicates the number of framing functions that have to be implemented within the FRA/DE-FRA blocks corresponding to each VNU. One may see that $3$ of the FRA/DE-FRA blocks must implement two framing functions, while the other $4$ FRA/DE-FRA blocks implement only one framing function.

\subsection{Full layers architecture}\label{subsec:full_layers_arch}

A different possibility to increase throughput is to increase the hardware parallelism, by including several non-overlapping rows of the base matrix in one decoding layer. For instance, for the base matrix in Fig.~\ref{fig:base_matrix_regular}, we may consider $\rpl=4$ consecutive rows per decoding layer, thus the number of decoding layers is $L=3$. In this case, each column of the base matrix has one (and only one) non-zero entry in each decoding layer; such a decoding layer is referred to as being {\em full}. Full layers correspond to the maximum hardware parallelism that can be exploited by layered architectures, but they also prevent the   pipelining of the data path.  
One possibility to implement a full-layer decoder is to use a similar architecture to the pipelined one, by removing the registers inserted after the VNU (since pipelining is incompatible with the use of full-layers), and updating the control unit. However, in such an architecture, read/write operations from/to the $\beta$\_memory would occur at the same memory location, corresponding to the current layer being processed $\ell$. This would require the use of asynchronous dual-port RAM to implement the $\beta$\_memory, which in general is known to be slower than synchronous dual-port RAM. The architecture proposed in this section, shown in Fig.~\ref{fig_MS_NS_FAID_1cc_short}, is aimed at avoiding the use of asynchronous RAM, while providing an effective way to benefit from the increased hardware parallelism enabled by the use of full layers.
We discuss below the main changes with respect to the pipelined architecture from the previous section, consisting of the $\alpha$\_memory and the barrel shifters blocks (the other blocks are the same as for the pipelined architecture), as well as a complete reorganization of the data path.  However, it can be easily verified that both architectures are logically equivalent, {\em i.e.}, they both implement the same decoding algorithm.

\subsubsection*{$\mathbf{\alpha}$\_memory} This memory is used to store the VN-messages for the current decoding layer (unlike the previous architecture, the AP-LLR values are not stored in memory). Since only one $\tilde{q}$-bit (unsaturated) VN-message is stored for each variable-node, this memory has exactly the same size as the $\tilde{\gamma}$\_memory used within the previous pipelined architecture. 
 VN-messages for current layer $\ell$ are read from the $\alpha$\_memory, then saturated or framed depending on the decoding kernel, and supplied to the corresponding CNUs. CN-messages computed by the CNUs are stored in the $\beta$\_memory (location corresponding to layer $\ell$), and also forwarded to the AP-LLR unit, through the DCP (decompress) and DE-FRA (de-framing) blocks, according to the CNU implementation (compressed or uncompressed) and the decoding kernel (MS of NS-FAID). The AP-LLR unit computes the sum of the incoming VN- and CN-messages, which corresponds to the AP-LLR value to be used at layer $\ell+1$ (since already updated by layer $\ell$). The AP-LLR value is forwarded to the VNU, through corresponding BS and PER blocks. Eventually, the VN-message for layer $\ell+1$ is computed as the difference between the incoming AP-LLR and the corresponding layer-$(\ell+1)$ CN-message computed at the previous iteration, the latter being read from the $\beta$\_memory.

\subsubsection*{PER\,/\,BS blocks} PER\_1\,/\,BS\_1 blocks  permute\,/\,shift the data read from the input buffer, according to the positions\,/\,values of the non-negative entries in the first decoding layer. Similarly to the BS\_R blocks in the pipelined architecture, the PER\_WR\,/\,BS\_WR blocks permute\,/\,shift the AP-LLR values, according to the {\em difference} between the  positions\,/\,values of the current layer's ($\ell$) non-negative entries  and those of the next layer ($\ell+1$). This way, VN-messages stored in the $\alpha$\_memory are already permuted and shifted for the subsequent decoding layer. Finally, PER\_$L$\,/\,BS\_$L$ blocks  permute\,/\,shift the hard decision bits (sign of AP-LLR values), according to the positions\,/\,values of the non-negative entries in the last decoding layer.



\section{Implementation Results}
\label{sec:nsfaid_implem_results}

This section reports implementation results, as well as the error correction performance of the implemented codes, in order to corroborate the analytic results obtained in Section~\ref{sec:de_optimization_ns_faids}. As mentioned in the previous section, both architectures are logically equivalent, thus they both yield the same decoding performance (assuming that they implement the same MS/NS-FAID decoding kernel), but they may have different performance in terms of area and throughput. 

\begin{figure*}[!t]
\centering
\subfigure[$(3,6)$-regular LDPC code]{\includegraphics[width=0.47\linewidth]{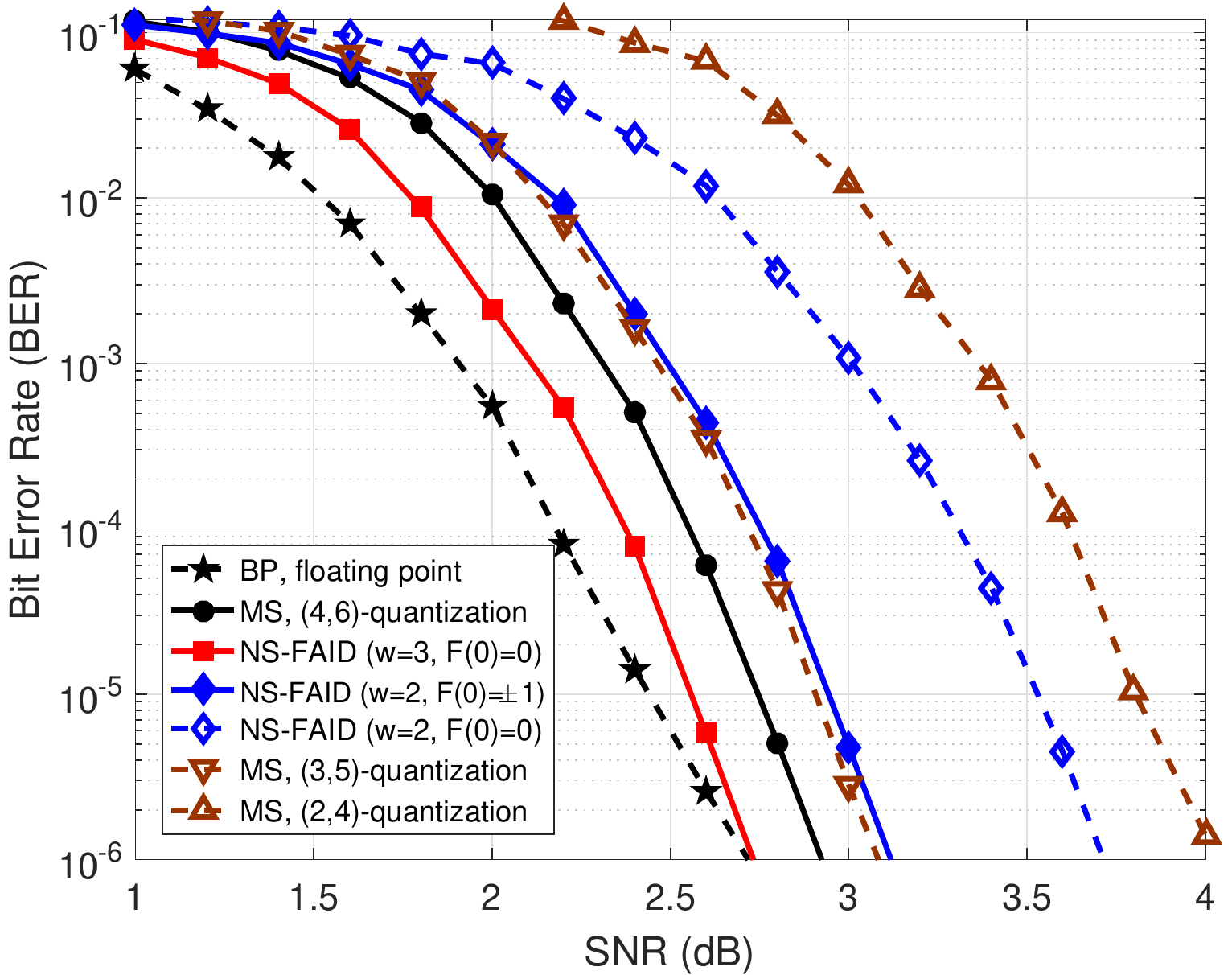}\label{fig:reg_nsfaid_ber}}%
\hfill\subfigure[WiMAX irregular LDPC code]{\includegraphics[width=0.47\linewidth]{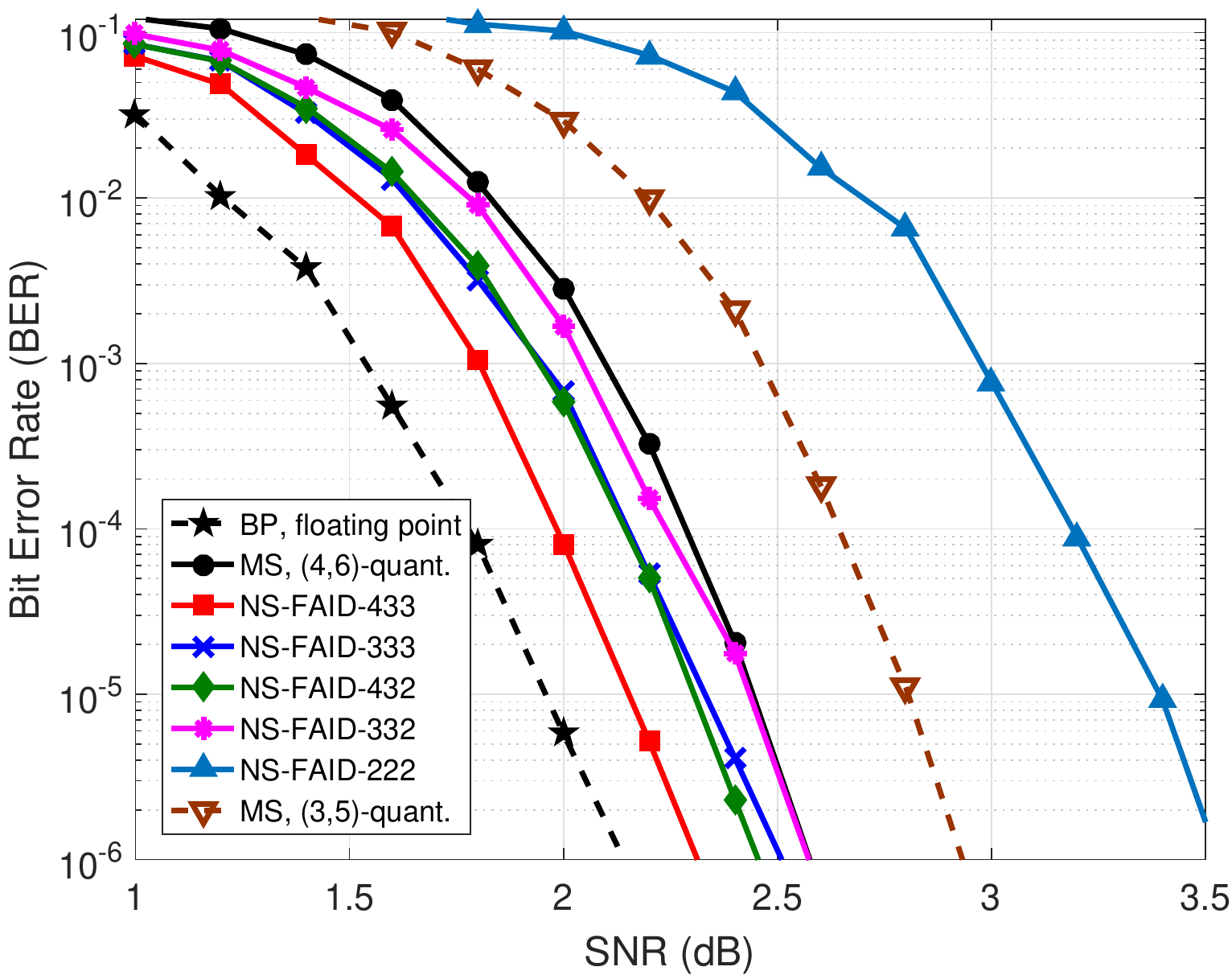}\label{fig:irreg_nsfaid_ber}}
\caption{BER performance of optimized regular and irregular NS-FAIDs}
\label{fig:reg_irreg_nsfaid_ber}
\end{figure*}

\begin{table*}[!t]
\caption{ASIC post-synthesis implementation results on 65nm-CMOS technology for $(3, 6)$ regular LDPC \vspace*{-2mm}}
\label{tab:reg_nsfaids_asic}
\fontsize{8pt}{9pt}\selectfont
\centering
\resizebox{\textwidth}{!}{\begin{tabular}{*{12}{@{}c@{}|}@{}c@{}}
\hline
\textbf{Variant} 
&\multicolumn{3}{@{}c@{}|@{}}{\textbf{pipelined.uncompressed}} &\multicolumn{3}{@{}c@{}|@{}}{\textbf{pipelined.compressed}} &\multicolumn{3}{@{}c@{}|@{}}{\textbf{full\_layers.uncompressed}} 
&\multicolumn{3}{@{}c@{}}{\textbf{full\_layers.compressed}}\\
\hline
\textbf{Decoder} & \,MS(4,6)\, &\,NS-FAID-3\, &\,NS-FAID-2\, &\,MS(4,6)\, &\,NS-FAID-3\, &\,NS-FAID-2\, &\,MS(4,6)\, &\,NS-FAID-3\, &\,NS-FAID-2\, &\,MS(4,6)\, &\,NS-FAID-3\, &\,NS-FAID-2\\
\hline
\hline
\textbf{Max. Freq. (MHz)} &200 &222 &227 &175 &200 &208 &151 &172 &192 &125 &147 &172\\
\hline
\textbf{Throughput (Mbps)} &1075 &1193 &1220 &941 &1075 &1118 &3261 &3715 &4147 &2700 &3175 &3715\\
\hline
\textbf{Area $(\text{mm}^2)$} &0.45 &0.42 &0.38 &0.41 &0.38 &0.36 &0.80 &0.72 &0.68 &0.75 &0.67 &0.65\\
\hline\hline
\textbf{TAR $(\text{Mbps}/{\text{mm}}^2)$} &2389 &2840 &3210 &2295 &2828 &3105 &4076 &5159 &6098 &3600 &4738 &5715\\
\textbf{$\pm{\%}$ w.r.t. MS(4,6)} & {0} & {+18.88} & {+34.37} & {0} & {+23.22} & {+35.29} & {0} & {+26.57} & {+49.61} & {0} & {+31.61} & {+58.75} \\
\hline
\end{tabular}}
\end{table*}

\subsection{Regular LDPC Codes}

We consider the $(3,6)$-regular QC-LDPC code with base matrix $B$ of size $R\times C = 12 \times 24$, shown in Fig.~\ref{fig:base_matrix_regular}. The expansion factor $z=54$, thus the codeword length is $N=zC=1296$ bits. The base matrix can be divided in either $L=12$ decoding layers ($\rpl=1)$, for the pipelined architecture, or $L=3$ horizontal decoding layers ($\rpl=4$), for the full-layer architecture. 

Fig. \ref{fig:reg_nsfaid_ber} shows the Bit-Error Rate (BER) performance of the MS decoder with quantization parameters $(q, \tilde{q}) = (4,6)$, as well as $q=4$-bit NS-FAIDs with $w=2$ and $w=3$ (framing functions $F$ corresponding to $w$ and $F(0)$ values in the legend are those from Table~\ref{tab:best_reg_nsfaid}). Binary input AWGN channel model is considered, with $20$ decoding iterations. It can be seen that the simulation results corroborate the analytic results from Section~\ref{subsec:opt_reg_nsfaid}, in terms of SNR gain\,/\,loss provided by NS-FAIDs, as compared to MS. For comparison purposes, we have further included simulations results for 
the floating-point Belief Propagation (BP) decoder \cite{savin2014ldpc}, as well as the MS decoder with $(3, 5)$ and $(2, 4)$-quantization.

ASIC post-synthesis implementation results on 65nm-CMOS technology are shown in Table~\ref{tab:reg_nsfaids_asic}, for the  MS$(4,6)$ decoder  and the NS-FAIDs  with $(w = 3, F(0)=0)$ and  $(w = 2, F(0)=\pm1)$, indicated in the table as NS-FAID-3 and NS-FAID-2, respectively. The first (Variant) row in Table~\ref{tab:reg_nsfaids_asic} indicates the architecture (pipelined or full layers) and the CNU type (compressed or uncompressed). We also note that for the NS-FAID-2, the assumption that $0$ is mapped to either $-1$ or $+1$, with equal probability, is only needed for theoretical analysis (the symmetry of the decoder allows reducing the analysis to the all-zero codeword). However, in practical situations one may always map $0$ to $+1$, since random codewords are transmitted (for instance, in telecommunications systems, pseudo-randomness of the transmitted data is ensured by a scrambling mechanism).

Throughput reported in Table~\ref{tab:reg_nsfaids_asic} is given by the formula:
\begin{equation}\label{eq:throughput}
\text{Throughput} = \frac{N \times f_{\text{max}}}{\delta + L \times n_\text{iter}},
\end{equation} 
where $f_{\text{max}}$ is the maximum operating frequency (post synthesis), $n_\text{iter}$ is the number of decoding iterations (set to $20$), and $\delta=1$ for the pipelined architecture or $\delta=0$ for the full layers architecture. To keep the throughput comparison on an equal basis, we further define the {\em Throughput to Area Ratio} metric TAR = Throughput / Area (Mbps/mm$^2$).  

While the NS-FAID-3 decoder  outperforms the baseline MS($4,6$) decoder by $0.19$\,dB at $\text{BER}=10^{-5}$ (Fig.~\ref{fig:reg_nsfaid_ber}), it can be seen from Table~\ref{tab:reg_nsfaids_asic} that it also exhibits a TAR improvement between $18.88\%$ and  $31.61\%$, depending on the hardware architecture and CNU type. As predicted, the NS-FAID-2 decoder exhibit a performance loss of $0.21$\,dB compared to MS($4,6$), but yields a significant TAR improvement, by $34.37\%$ to $58.75\%$.

\subsection{Irregular LDPC Codes}

We consider the irregular WiMAX QC-LDPC code of rate $1/2$, with base matrix of size $R\times C = 12 \times 24$ \cite{wimax}. The expansion factor $z=96$, thus resulting in a codeword length $N=zC=2304$ bits. The pipelined architecture from Section~\ref{subsec:pipelined_arch} is implemented, with $\rpl=1$ row per decoding layer, after reordering the rows of the base matrix, such that any two consecutive rows do not overlap \cite{zhang2015full}. Note that the full layers architecture does not apply to irregular WiMAX LDPC codes, since it is not possible to group the rows of the base matrix in full decoding layers. 

BER results for the MS$(4,6)$ decoder  and the NS-FAID-$w_1w_2w_3$ decoders from Table~\ref{tradeoff} are shown in Fig.~\ref{fig:irreg_nsfaid_ber}, while ASIC post-synthesis implementation results on 65nm-CMOS technology are shown in Table~\ref{tab:irreg_nsfaids_asic}. The throughput reported is computed using Eq.~(\ref{eq:throughput}). TAR results and corresponding gain/loss ($+/-$) with respect to the MS$(4,6)$ decoder are reported on the last row. The NS-FAID-433 and NS-FAID-432 decoders outperform the MS decoder by $0.3$\,dB and
$0.15$\,dB (at BER\,$=\!10^{-5}$), respectively, at the price of a small degradation of the TAR. NS-FAIDs-333 improves the BER performance by $0.12$\,dB, with TAR improvement by $13.51\%$ to $16.39\%$, depending on the CNU type (compressed or uncompressed). NS-FAIDs-332 exhibits similar BER performance, with TAR improvement by $13.51\%$ to $19.30\%$. The NS-FAID-222 decoder yields the most significant TAR improvement (up to $42.09\%$), but this comes at the price of a significant BER degradation by $\approx 1$\,dB as estimated in Section~\ref{subsec:opt_irreg_nsfaid}. 

To further emphasize the high-throughput characteristic of the proposed architecture, the irregular NS-FAID-332 decoder is further compared with other state of the art implementations of WiMAX decoders in Table~\ref{tab:WiMAX_comparison}. We also
report the TAR and Normalized TAR (NTAR) metrics, so as to keep the throughput comparison on an equal basis
with respect to technology, area, and number of iterations. To scale throughput and area to $65$nm, we use scale factors (technology\_size/$65$) and ($65$/technology\_size)$^2$, respectively, as suggested in \cite{hauser2008mosfet}. The computation of the TAR and NTAR metrics is detailed in the footnote to Table~\ref{tab:WiMAX_comparison}. Note that for all the reported implementations, the achieved throughput is inversely proportional to the number of iterations, hence the NTAR metric corresponds to the TAR value assuming that only one decoding iteration is performed. We mention that the decoders proposed in \cite{zhang2015full, kanchetla2016multi} are reconfigurable decoders that support the IEEE 802.16e (WiMAX) and and the IEEE 802.11n (WiFi) wireless standards. The reported throughput is the maximum achievable coded throughput for the $(1152, 2304)$ WiMAX code, with either $10$ or $5$ decoding iterations. From Table~\ref{tab:WiMAX_comparison} it can be seen that the proposed irregular NS-FAID compares favorably with state of the art implementations, yielding a NTAR value of $45.86$\,Gbps/mm$^2$/iteration.

\begin{table*}[!t]
\caption{ASIC post-synthesis implementation results on 65nm-CMOS technology for WiMAX LDPC\vspace*{-2mm}}
\label{tab:irreg_nsfaids_asic}
\fontsize{8pt}{9pt}\selectfont
\centering
\resizebox{\textwidth}{!}{\begin{tabular}{*{12}{@{}c@{}|}@{}c@{}}
\hline
\textbf{Variant} 
&\multicolumn{6}{@{}c@{}|@{}}{\textbf{pipelined.uncompressed}}&\multicolumn{6}{@{}c@{}}{\textbf{pipelined.compressed}}\\
\hline 
\multirow{2}{23mm}{\centering\begin{tabular}{@{}c@{}}\textbf{Decoder}\end{tabular}}  &\,MS(4,6)\, &\,NS-FAID\, &\,NS-FAID\, &\,NS-FAID\, &\,NS-FAID\, &\,NS-FAID\, &\,MS(4,6)\, &\,NS-FAID\, &\,NS-FAID\, &\,NS-FAID\, &\,NS-FAID\, &\,NS-FAID\,\\
&  &433 &432 &333 &332 &222 &  &433 &432 &333 &332 &222\\
\hline
\hline
\textbf{Max. Freq. (MHz)} &175 &172 &178 &192 &192 &200 &161 &156 &161 &178 &178 &200\\
\hline
\textbf{Throughput (Mbps)} &1673 &1644 &1701 &1835 &1835 &1912 &1539 &1491 &1539 &1701 &1701 &1912\\
\hline
\textbf{Area $(\text{mm}^2)$} &0.87 &0.88 &0.90 &0.82 &0.80 &0.70 &0.77 &0.79 &0.79 &0.75 &0.75 &0.72\\
\hline\hline
\textbf{TAR $(\text{Mbps}/{\text{mm}}^2)$} &1922 &1868 &1890 &2237 &2293 &2731 &1998 &1887 &1948 &2268 &2268 &2655\\
\textbf{$\pm{\%}$ w.r.t. MS(4,6)} 
&{0} &{-2.81}&{-1.66}&{+16.39}&{+19.30}&{+42.09}&{0}&{-5.56}&{-2.50}&{+13.51}&{+13.51}&{+32.88}\\
\hline
\end{tabular}}
\end{table*}

\begin{table*}[!t]
\centering
\caption{Comparison between the proposed NS-FAID and state of the art implementations for the WiMAX QC-LDPC code}
\label{tab:WiMAX_comparison}
\fontsize{8pt}{9pt}\selectfont
\centering
\begin{tabular}{@{\ }c@{\ }|*{5}{@{\ \,}c@{\ \,}|}@{\ }c@{\ }}
\multirow{2}{23mm}{\centering\begin{tabular}{@{}c@{}}\textbf{Decoders}\end{tabular}}  
&\multirow{2}{16mm}{\centering\begin{tabular}{@{}c@{}}\textbf{K. Zhang'09}\\ \textbf{\cite{zhang2009high}}\end{tabular}}
&\multirow{2}{16mm}{\centering\begin{tabular}{@{}c@{}}\textbf{B. Xiang'11}\\ \textbf{\cite{xiang2011847}}\end{tabular}}
&\multirow{2}{17mm}{\centering\begin{tabular}{@{}c@{}}\textbf{T. Heidari'13}\\ \textbf{\cite{heidari2013design}}\end{tabular}}
&\multirow{2}{17mm}{\centering\begin{tabular}{@{}c@{}}\textbf{W. Zhang'15}\\ \textbf{\cite{zhang2015full}}\end{tabular}}
&\multirow{2}{20mm}{\centering\begin{tabular}{@{}c@{}}\textbf{K. Kanchetla'16}\\ \textbf{\cite{kanchetla2016multi}}\end{tabular}}
&\multirow{2}{17mm}{\centering\begin{tabular}{@{}c@{}}\textbf{This work}\\ \textbf{NS-FAID-332}\end{tabular}}\\
& & & & &\\
\hline
\textbf{Code length}  &2304 &576-2304 &2304 &576-2304$^{(\dagger)}$ &576-2304$^{(\dagger)}$ &2304\\
\hline
\textbf{Technology (nm)}   &90 &130  &130  &40  &90 &65\\
\hline
\textbf{Frequency (MHz)}     &950 &214  &100  &290  &149 &192\\
\hline
\textbf{Iterations}   &10  &10  &10  &10  &5  &20\\
\hline
\textbf{Throughput (Mbps)} &2200 &955  &183  &2227  &955  &1835\\
\textbf{Tput scaled to 65nm (Mbps)} &3036  &1910  &366  &1370  &1318 &1835\\
\hline
\textbf{Area (mm$^2$)} &2.90$^{(\ast)}$  &3.03$^{(\ast)}$  &6.90$^{(\ast \ast)}$  &2.26$^{(\ast)}$  &11.42$^{(\ast)}$ &0.80$^{(\ast)}$\\
\textbf{Area scaled to 65nm (mm$^2$)} &1.51$^{(\ast)}$  &0.76$^{(\ast)}$ &1.73$^{(\ast \ast)}$  &5.97$^{(\ast)}$ &5.94$^{(\ast)}$ &0.80$^{(\ast)}$\\
\hline
\textbf{TAR (Mbps/mm$^2$)} &2011 &2513  &212  &229  &222 &2293\\
\hline
\textbf{NTAR (Mbps/mm$^2$/iter)} &20110 &25130  &2120 &2290 &1110 &45860\\
\hline
\multicolumn{6}{l}{$^{(\dagger)}$ support both WiMAX and Wi-Fi standards} \\
\multicolumn{6}{l}{$^{(\ast)}$ only core area is reported} \\
\multicolumn{6}{l}{$^{(\ast \ast)}$ total chip area is reported} \\
\multicolumn{6}{l}{TAR =  (Throughput scaled to 65nm) / (Area scaled to 65nm)} \\
\multicolumn{6}{l}{NTAR =  TAR $\times$ Iterations}
\end{tabular}
\end{table*}

\section{Conclusion}
\label{ns_faids:sec:Conclusion}
In this paper, we first introduced the new framework of Non-Surjective FAIDs, which allows trading off decoding performance for hardware complexity reductions. NS-FAIDs have been optimized by density evolution and shown to provide significant memory size reductions, with similar or event better decoding performance, as compared to the MS decoder. Then, two hardware architectures have been presented, making use of either pipelining or increased hardware parallelism in order to increase throughput. Both MS and NS-FAID decoding kernels have been integrated into each of the two proposed architectures, and compared in terms of area and throughput. ASIC post synthesis implementation results demonstrated the effectiveness of the NS-FAID approach in yielding  significant improvements in terms of area and throughput, as compared to the MS decoder, with even better or only slightly degraded decoding performance.


\section*{Acknowledgment}
The authors acknowledge support from the European H2020 Work Programme, project Flex5Gware, and the Franco-Romanian (ANR-UEFISCDI) Joint Research Programme ``Blanc-2013'', project DIAMOND.


\bibliographystyle{IEEEtran}
\bibliography{biblio}

\vspace*{-7mm}
\begin{IEEEbiography}[{\includegraphics[width=1in,height=1.25in,clip,keepaspectratio]{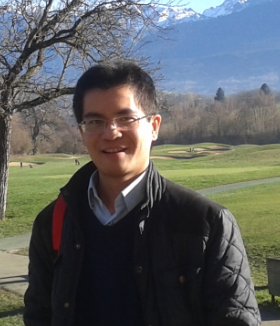}}]{Truong Nguyen-Ly} received his B.S. and M.S. degrees in Electronic and Telecommunication engineering from Ho Chi Minh City University of Technology (HCMUT), Vietnam, in 2010 and 2012, respectively. From 2010 to 2014, he was a lecturer at Faculty of Electrical and Electronic Engineering, HCMUT, Vietnam. He is currently working toward the Ph.D. degree 
in Telecommunications Engineering at the Broadband Wireless Systems Laboratory, CEA-LETI, MINATEC Campus, and ETIS ENSEA/UCP/CNRS UMR-8051, France. His research interests include error-correction coding, analysis and implementation of LDPC decoder architectures on FPGA/ASIC platform, and speech processing.
\end{IEEEbiography}

\begin{IEEEbiography}[{\includegraphics[width=1in,height=1.25in,clip,keepaspectratio]{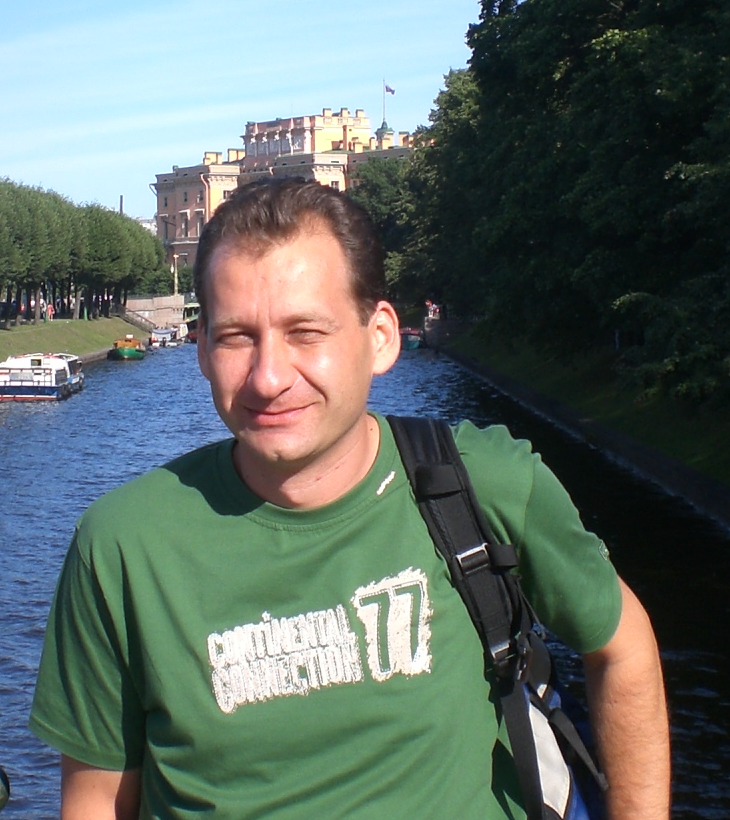}}]{Valentin Savin}
received his Master's Degree in Mathematics from the ``Ecole Normale Supérieure'' of Lyon in 1997, and his PhD in Mathematics from J. Fourier Institute, Grenoble, in October 2001. He also holds a Master's Degree in Cryptography, Security and Coding Theory from the University of Grenoble 1. Since 2005, he has been with the Digital Communications Laboratory of CEA-LETI, first as a two-year postdoctoral fellow, and then as a research engineer. Since 2016, he has been appointed CEA Senior Expert on information and coding theory. Over the last years, he has been working on the design of low-complexity decoding algorithms for LDPC and Polar codes, and on the analysis and the optimization of LDPC codes for physical and upper-layers applications.
He has published more than 70 papers in international journals and conference proceedings, holds 10 patents, and is currently participating in or coordinating several French and European research projects in ICT.
\end{IEEEbiography}

\begin{IEEEbiography}[{\includegraphics[width=1in,height=1.25in,clip,keepaspectratio]{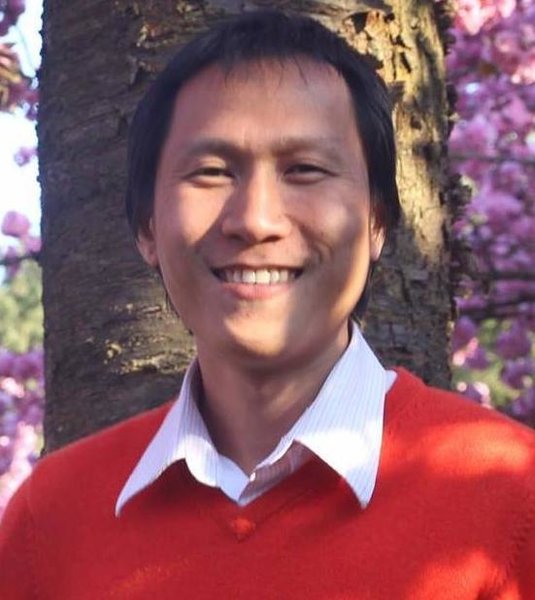}}]{Khoa Le}
received his bachelor and Master of science degree in Electronics and Telecommunication Engineering from Ho Chi Minh City University of Technology (HCMUT), Vietnam in 2010 and 2012, respectively.
He is working toward the Ph.D degree at ETIS Laboratory, ENSEA, University of Cergy-Pontoise, CNRS UMR-8051, France.
His research interests are in error correcting code algorithms, analysis and their implementations in FPGA/ASIC.
\vspace*{-7mm}
\end{IEEEbiography}


\begin{IEEEbiography}[{\includegraphics[width=1in,height=1.25in,clip,keepaspectratio]{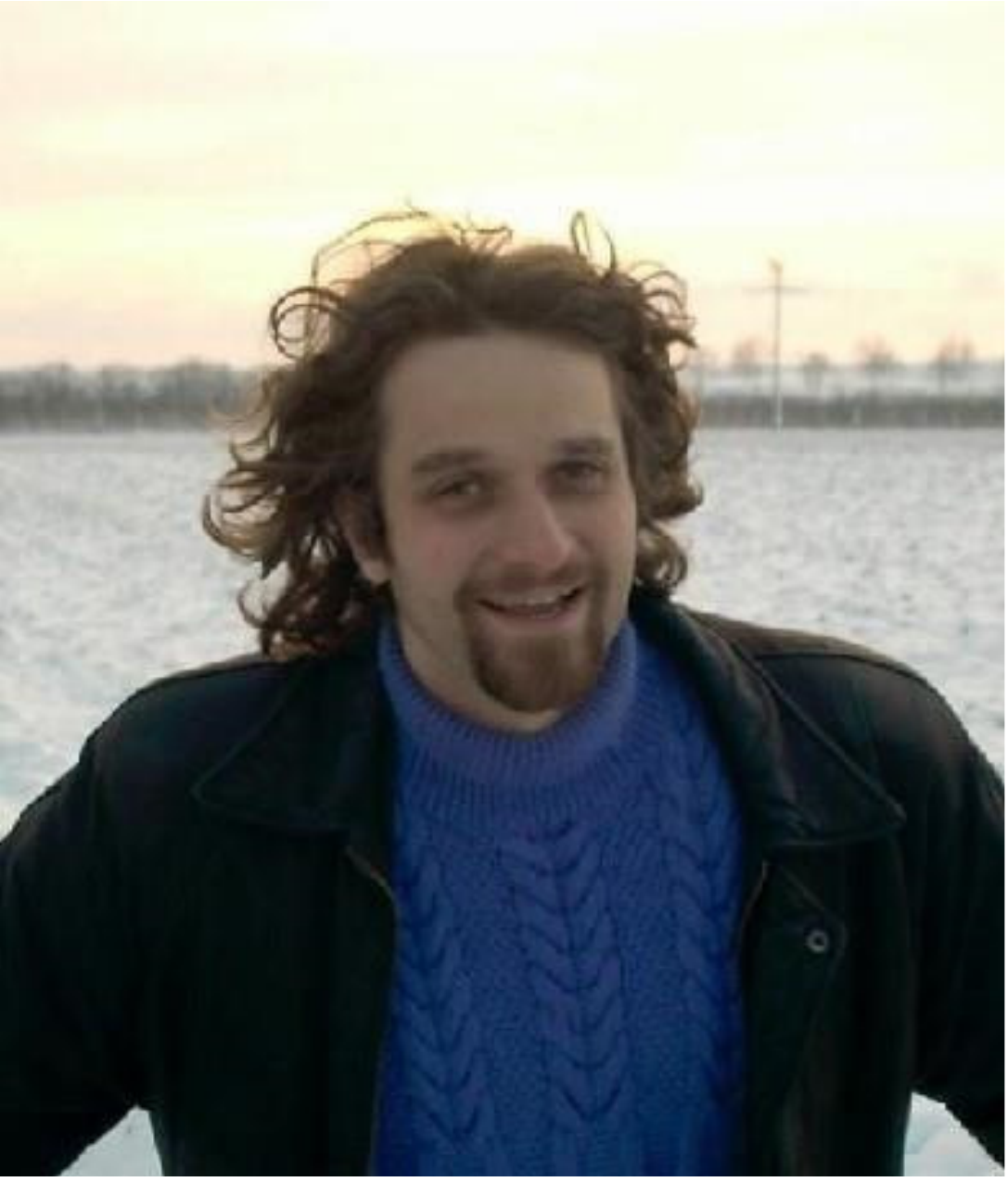}}]{David
Declercq} was born in June 1971. He graduated his PhD in Statistical
Signal Processing 1998, from the University of Cergy-Pontoise, France.
He is currently full professor at the ENSEA in Cergy-Pontoise.
He is the general secretary of the National GRETSI
association, and Senior member of the IEEE. He has held
the junior position at the ``Institut Universitaire de France'' from 2009 to 2014.
His research topics lie in digital communications and error-correction
coding theory. He worked several years on the particular family of LDPC
codes, both from the code and decoder design aspects.
Since 2003, he developed a strong expertise on non-binary LDPC codes
and decoders in high order Galois fields GF($q$). A large part of
his research projects are related to non-binary LDPC codes. He mainly
investigated two aspects: (i) the design of GF($q$) LDPC codes for short
and moderate lengths, and (ii) the simplification of the iterative
decoders for GF($q$) LDPC codes with complexity/performance tradeoff
constraints.
David Declercq published more than 40 papers in major journals
(IEEE-Trans. Commun., IEEE-Trans. Inf. Theo., Commun. Letters, EURASIP
JWCN), and more than 120 papers in major conferences in Information
Theory and Signal Processing.
\vspace*{-7mm}
\end{IEEEbiography}

\begin{IEEEbiography}[{\includegraphics[width=1in,height=1.25in,clip,keepaspectratio]{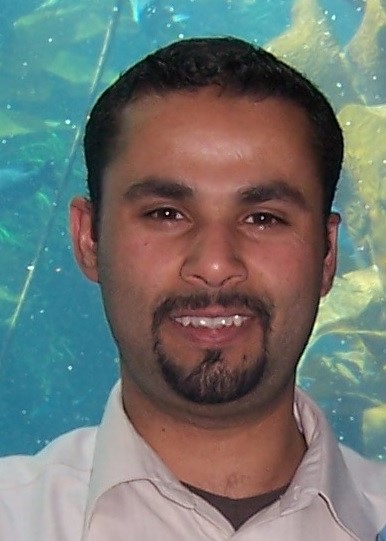}}]{Fakhreddine Ghaffari} received the Electrical Engineering and Master degrees from the National School of Electrical Engineering (ENIS, Tunisia), in 2001 and 2002, respectively. He received the Ph.D degree in electronics and electrical engineering from the University of Sophia Antipolis, France in 2006. He is currently an Associate Professor at the University of Cergy Pontoise, France. His research interests include VLSI design and implementation of reliable digital architectures for wireless communication applications in ASIC/FPGA platforms and the study of mitigating transient faults from algorithmic and implementation perspectives for high-throughput applications.
\vspace*{-7mm}
\end{IEEEbiography}

\begin{IEEEbiography}[{\includegraphics[width=1in,height=1.25in,clip,keepaspectratio]{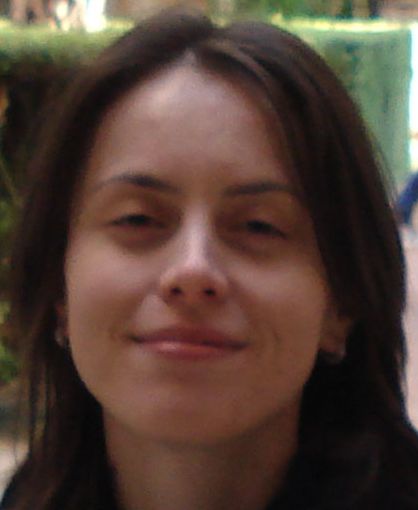}}]{Oana Boncalo}
received her B.Sc.and Ph.D. degree in Computer Engineering from the University Politehnica Timisoara, Romania, in 2006 and 2009 respectively. She is currently an Associate Professor at University Politehnica Timisoara.
She has published over 50 research papers in topics related to digital design. Her research interests include computer arithmetic, LDPC decoder architectures, digital design and reliability estimation and evaluation.
\vspace*{-7mm}
\end{IEEEbiography}


\begin{IEEEbiographynophoto}{Note concerning prior work:}
Preliminary version of part of this work has been previously published in \cite{nguyen2016non, nguyen2017high}. In this paper, the previous definition and density-evolution analysis of NS-FAIDs \cite{nguyen2016non} is extended to framing functions with $F(0) =\pm\lambda$, such as to cover a larger class of decoders, which is shown to significantly improve the decoding performance in case that the exchanged messages are quantized on a small number of bits (e.g., $2$ bits per exchanged message). Optimization results presented in Section~\ref{sec:de_optimization_ns_faids} are new, and they report on the optimization of regular and irregular NS-FAIDs, by taking into account the proposed extension. The hardware architectures proposed in \cite{nguyen2017high}  have been extended to cover the case of irregular NS-FAIDs. In addition, implementation results reported in this paper target an ASIC technology, which is more likely to reflect the benefits of the proposed NS-FAID approach in terms of throughput/area trade-off. All the implementation results reported in Section~\ref{sec:nsfaid_implem_results} (for both regular and irregular codes) are new.
\end{IEEEbiographynophoto}

\end{document}